\documentclass[preprint,authoryear,12pt]{elsarticle}

\usepackage{amssymb,amsmath}

\usepackage{lineno}

\bibpunct[, ]{(}{)}{;}{a}{,}{,}

\def\ssz{\scriptsize} 
\def\ds{\displaystyle} 
\def\ss{\scriptstyle}

\def\eps{\varepsilon}
\def\th{\theta}
 
\newcommand{\Frac}[2]{\ds\frac{\ds #1}{\ds #2}}

\newcommand{\onethird}{\mbox{$\frac{\ss 1}{\ss 3}\,$}}

\journal{ArXiv}

\begin{document}

\begin{frontmatter}

\addtolength\textheight{-24pt}



\title{The ecology of asexual pairwise interactions:\\ A generalized law of mass action}


\author{\vspace*{-1mm}
Fabio Dercole}

\address{Dept. of Electronics, Information, and Bioengineering, Politecnico di Milano, Italy\\
{\em address:} Via Ponzio 34/5, 20133 Milano, Italy\\
{\em ph[fax]:} +39 02 2399 3484 {\em[}3412{\em\hspace{0.5mm}];} {\em e-mail:} {\tt fabio.dercole@polimi.it}
\vspace*{-11mm}}

\begin{abstract}
A general procedure to formulate asexual (unstructured, deterministic) population dynamical models resulting from individual pairwise interactions is proposed.
Individuals are characterized by a continuous strategy that represents all their behavioral, morphological, and functional traits.
Populations group conspecific individuals with identical strategy and are measured by densities in space.
Species can be monomorphic, if only one strategy is present, or polymorphic otherwise.
The procedure highlights the structural properties fulfilled by the population per-capita growth rates.
In particular, the effect of perturbing a set of similar strategies is proportional to the product of the corresponding densities, with a proportionality coefficient that is density-dependent only through the total density.
This generalizes the law of mass action, which traditionally refers to the case in which the per-capita growth rates are linearly density-dependent and insensitive to joint strategy perturbations.
Being underpinned with individual strategies, the proposed procedure is most useful for evolutionary considerations, in the case strategies are inheritable.
The developed body of theory is exemplified on a Holling-type-II many-prey-one-predator system and on a model of a cannibalistic community.
\end{abstract}

\begin{keyword}
ecological modeling\sep
functional response\sep
individual strategy\sep
invasion fitness\sep
Lotka-Volterra\sep
mass action\sep
pairwise interactions\sep
phenotypic trait\sep
population dynamics.


\vspace*{-2mm}
\end{keyword}

\end{frontmatter}



\section{Introduction}
\label{sec:int}
A significant part of ecological modeling describes large (technically infinite) populations, measured by finite densities per unit of space and composed of asexual individuals interacting in pairs \citep[see, e.g.,][]{Gurney98,Thieme_03_PUP,Hastings_and_Gross_12_UCA}.
The underlying paradigm is that of the ``mass action,'' introduced in 1864 as a model for the kinetics of chemical reactions
\citep{Waage_and_Gulberg_1864,Lund_65_JCED}.
In the ecological context, individuals are thought as the molecules of a well mixed fluid and population densities as the concentrations of the chemical species.
Individuals can react by themselves (on a Poissonian basis) turning from one state into another---e.g. dying or switching from food handling to searching ---or at pairwise random encounters---e.g. between prey and predator.
The simultaneous encounter of more than two individuals is a higher-order process
(with a probability to occur in an infinitesimal time interval $\mathrm{d}t$ that vanishes at least as $\mathrm{d}t^2$)
and is disregarded (it gives no contribution in the limit $\mathrm{d}t\to 0$).

The {\it law of mass action} leads to Lotka-Volterra ecological models \citep{Lotka_20_JACS,Volterra26}, i.e., ODE models for the dynamics of the population densities characterized by density dependencies that are at most quadratic.
In other words, once factoring the population density, the remaining {\it per-capita} (or per-unit-of-density) growth rate is linearly density-dependent.
Think, e.g., of a monomorphic predator harvesting a polymorphic prey.
The model can be given in terms of equivalent chemical reactions, that translate into the ODEs describing the reactions' kinetics (see Box~1).
Prey individuals are characterized by a continuous strategy $x$, representing all their behavioral, morphological, and functional traits, and individuals with same strategy $x_i$ are grouped into the population with density $n_i$, $i=1,\hspace{-0.6mm}...,\hspace{-0.3mm}M$.
Predators are all identical.
They could also be described in terms of a strategy, but this is left implicit to focus on prey polymorphism.

Nonlinear per-capita growth rates can be obtained by structuring the populations into behavioral states, setting mass action laws for the density fluxes among the different states, and using time-scale separation arguments to force the fast behavioral dynamics at equilibrium \citep[see e.g.][]{Ruxton_et_al_TPB_92}.
The famous Holling type II functional response for predation can be derived along this line (see Box~2).

A nonlinear density dependence in the per-capita growth rate of an unstructured population---hereafter ``nonlinear fitness'' \citep{Metz92}---describes the fact that individuals encounter other individuals who are at the same time involved in other encounters or activities.
As a result, not all the encounters trigger the corresponding reaction.
Prey, for example, do not react with all encountered predators, as some of them are handling other prey.
This becomes evident when the predator population is structured into behavioral states (Box~2), as prey react only with searching predators.

In this paper I propose a procedure to directly build nonlinear fitnesses, explicitly taking into account that pairwise interactions can depend on the concomitant activities of the encountered individuals.
The focus is on a given species---e.g. the prey as in Box~1---which is thought to be polymorphic, with individuals grouped according to the value of a continuous strategy $x$.
The {\it focus} species may interact with other species, the strategies of which are however left implicit.
The target is to define the so-called {\it fitness generating function} \citep[or $g$-function,][]{Vincent05} for the focus species, namely a function of a ``virtual'' strategy $x'$ that gives for $x'=x_i$ the per-capita growth rate of population $i$, $i=1,\hspace{-0.6mm}...,\hspace{-0.3mm}M$
(technically, the $g$-function gives the fitness of the strategy $x'$ at zero density in an environment where the strategies $x_1,x_2,\hspace{-0.6mm}...,x_M$, as well as the other species, are present at nonzero densities).
The $g$-function takes as arguments, other than the virtual strategy $x'$, a set of ``environmental variables'' which describe how the strategy $x'$ interacts with all the biotic and abiotic components of its environment.
The core of the approach is in the definition of these environmental variables.
In particular, the intra-specific interactions between the strategy $x'$ and the {\it resident} strategies $x_1,x_2,\hspace{-0.6mm}...,x_M$ are built through an iterative procedure that, at each step, integrates over the strategy space suitable interacting terms, which, in turn, may depend on the integrals computed in the previous steps.
In this way, the interactions that do not depend on the concurrent activities of the interacting conspecifics only need one step, while those depending on the others' activities up to a certain level of nesting do require more steps.

The so constructed $g$-function has built-in all the structural properties one should expect to hold, for consistency, when some of the conspecific morphs are similar or vanishing.
E.g., if $x_1\hspace{-0.5mm}=x_2$ one should be able to rewrite a reduced $g$-function where only one of the two strategies is present with density $n_1+\hspace{0.5mm}n_2$.
And a similar property holds for the $g$-function derivatives w.r.t.\hspace{-0.5mm} the resident strategies:
the effect (on the growth of the virtual strategy $x'$) of perturbing a resident strategy is proportional to the corresponding density through a coefficient that, when all strategies are equal, can be density-dependent only through the total density of the species.
This is again the law of mass-action (as it can be verified in Box~1), using which, however, the proportionality coefficient is density-independent.
The proposed procedure therefore generalizes the law.
Specifically, it shows that the effect of a higher-order perturbation w.r.t.\hspace{-0.5mm} a subset of the resident strategies is given by a sum of terms, each proportional to the product of the corresponding densities, with each density up to the power of the corresponding differentiation order.
And the proportionality coefficients depend on the total density of the species when all strategies are equal.
Note, in particular, that mixed derivatives, w.r.t.\hspace{-0.5mm} different resident strategies, may have effect, though this is not the case when strict mass-action is used (Box~1), resulting only in a linear density effect in the pure fitness derivatives of any order.

Although quite technical, the generalized mass-action property has a great impact in the theoretical analysis, in particular in studying the competition between similar strategies.
This is typically done by expanding the $g$-function of the focus species around a reference strategy $x$, so the $g$-derivatives w.r.t.\hspace{-0.5mm} the resident strategies that appear in the expansion are evaluated with all resident strategies at $x$, and the property can be usefully exploited.
Specifically, it is possible to express the $g$-derivatives in the expansion in terms of the derivatives of the monomorphic $g$-function---the $g$-function for $M=1$.
This is particularly interesting because it means that the competition between the reference strategy and similar (potentially invading) ones can be studied by looking at reference strategy alone.
When we go to evolutionary considerations \citep{Dercole_and_Rinaldi_08_PUP}, the expansion of the dimorphic fitness ($M=2$) w.r.t.\hspace{-0.5mm} $(x_1,x_2)$ around $(x,x)$ can be again expressed, thanks to the generalized mass action, in terms of the monomorphic fitness.
The evolutionary dynamics of the focus species locally to an evolutionary branching \citep{Geritz97,Geritz98} can then be studied in terms of the geometry of the monomorphic fitness.

The paper is organized as follows. 
In Sect.~\ref{sec:met} the general procedure to build the $g$-function for the focus species is presented (Subsects.~\ref{ssec:not} and~\ref{ssec:proc}) and its structural properties highlighted (Subsects.~\ref{ssec:prop} and~\ref{ssec:how}).
Boxes~3 and~4 exemplify the procedure for the $M$-prey-one-predator model in Box~2 and for an evolutionary model of cannibalism taken from the literature \citep{Dercole_and_Rinaldi_02_TPB,Dercole_03_JMB}.
All the computations are performed in a {\it Mathematica} script accompanying the paper (as Supplementary Material), that can be easily adapted for use with any ecological models.
Sect.~\ref{sec:dis} discusses the obtained results and, in particular, their relevance in the opening of new theoretical research directions.

\section{Methods}
\label{sec:met}

\subsection{Basic notation and assumptions}
\label{ssec:not}
Consider an $M$-morphic species---the focus species---characterized by strategies $x_1,x_2,\hspace{-0.6mm}...,x_M$ and population densities $n_1(t),n_2(t),\hspace{-0.6mm}...,n_M(t)$ at time $t$.
Individuals can interact in pairs with other conspecifics, or with individuals belonging to other species, the latter described by a set of $P$ populations (some of which may be conspecific) with densities $N_1(t),N_2(t),\hspace{-0.6mm}...,N_P(t)$.
The ecological dynamics of the population densities are modeled in continuous time by a set of ODEs of the form
\begin{linenomath}
\begin{subequations}
\label{eq:eco1}
\begin{eqnarray}
\label{eq:nidot}
\dot{n}_i & = & n_i\,g_M(n_{1:M},N_{1:P},x_{1:M},x_i,t),\quad i=1,\hspace{-0.6mm}...,\hspace{-0.3mm}M,\\
\label{eq:Njdot}
\dot{N}_{1:P} & = & F_M(n_{1:M},N_{1:P},x_{1:M},t),
\end{eqnarray}
\end{subequations}
\end{linenomath}
where, hereafter, $n_{1:M}$, $N_{1:P}$, and $x_{1:M}$ respectively stand for $n_1,\hspace{-0.6mm}...,n_M$, $N_1,\hspace{-0.6mm}...,N_P$, and $x_1,\hspace{-0.6mm}...,x_M$,
\begin{linenomath}
$$
g_M(n_{1:M},N_{1:P},x_{1:M},x'\hspace{-0.5mm},t)
$$
\end{linenomath}
is the $M$-morphic $g$-function associated with the focus species \citep{Vincent05}, and
\begin{linenomath}
$$
F_M(n_{1:M},N_{1:P},x_{1:M},t)
$$
\end{linenomath}
is a vector of $P$ functions collecting the population growth rates of the populations $N_1,\hspace{-0.6mm}...,N_P$ of different species
(note that $F_{M,p}/N_p$ is the per-capita growth rate of population $N_p$, $p=1,\hspace{-0.6mm}...,\hspace{-0.3mm}P$).

Namely, $g_M$ is the per-capita growth rate of a (virtual) strategy $x'$ with zero density in an environment set by the resident strategies $x_1,x_2,\hspace{-0.6mm}...,x_M$, with densities $n_1,n_2,\hspace{-0.6mm}...,n_M$, together with the populations $N_1,\hspace{-0.6mm}...,N_P$ of different species and the abiotic environmental conditions at time $t$.
Note the list of arguments in functions $g_M$ and $F_M$.
First the dynamical variables $n_{1:M}$ and $N_{1:P}$, then the strategies $x_{1:M}$ and $x'$, which play the role of parameters in model \eqref{eq:eco1}, and last the time $t$ that takes possible time-dependencies of the abiotic environment into account (e.g. seasonalities and climatic fluctuations).

Let me define the density distribution $\nu$ of the resident strategies
\begin{linenomath}
\begin{equation}
\label{eq:nu}
\nu(x,x_{1:M},n_{1:M}):=\sum_{j=1}^{M}n_j\delta(x-x_j),
\end{equation}
\end{linenomath}
where $x$ is a variable spanning the strategy space and $\delta(x-x_j)$ is the Dirac peak at $x=x_j$.
Note that $\nu$ does not explicitly depend on time, though it changes in time due to the dynamics of the densities $n_1\hspace{-0.2mm},\hspace{-0.2mm}n_2,\hspace{-0.5mm}...,n_M$.

As explicitly noted in \citet{Meszena05a}, a contribution that I will further discuss in Sect.~\ref{sec:dis}, the $g$-function operates on $\nu$
(other than on $(x'\hspace{-0.5mm},N_{1:P},t)$).
This is the key fact ensuring that $g_M$ satisfies all natural consistency properties, namely the fact that if the virtual strategy $x'$ interacts with strategy $x_i$, the same interaction then occurs with strategies $x_j$, $j\neq i$, and if $x_i=x_j$ then the sum $n_i+n_j$ matters in lieu of $n_i$ and $n_j$ independently.
Instead of having $\nu$ as a direct argument of the $g$-function \citep[as in][]{Meszena05a}---which requires the definition of the functional derivative w.r.t.\hspace{-0.5mm} a distribution---I use the notion of {\it environmental feedback} \citep[introduced by][]{Mylius_and_Diekmann_95_OI,Metz_and_Gyllenberg_01_PRSB} to define the function $g_M$, for $M=1,2,\hspace{-0.6mm}...$, through the application of an $M$-independent function $g$ that operates on the virtual strategy $x'$ and on a set of relevant ``environmental quantities.''
In formulas, I set
\begin{subequations}
\begin{linenomath}
\begin{multline}
\label{eq:G}
g_M(n_{1:M},N_{1:P},x_{1:M},x'\hspace{-0.5mm},t):=\\
g(x'\hspace{-0.5mm},E_n(x'\hspace{-0.5mm},x_{1:M},n_{1:M},N_{1:P},e(t)),E_N(x_{1:M},n_{1:M},N_{1:P},e(t)),N_{1:P},e(t)),
\end{multline}
\end{linenomath}
where
\begin{itemize}
\vspace*{-2mm}\item[--]
$E_n(x'\hspace{-0.5mm},x_{1:M},n_{1:M},N_{1:P},e)$ is a set of functions of the strategy $x'$ that characterize the interactions of the $x'$-strategist with its conspecifics
(these functions may depend on $N_{1:P}$ if the interaction extends to some of the other species, e.g. predation in Box~4);
\vspace*{-2mm}\item[--]
$E_N(x_{1:M},n_{1:M},N_{1:P},e)$ is a set of functions describing the interactions of the individuals of populations $N_1,\hspace{-0.6mm}...,N_P$ with the resident strategies $x_1,x_2,\hspace{-0.6mm}...,x_M$
(these interactions may affect those with the $x'$-strategists);
\vspace*{-6.5mm}\item[--]
$e(t)$ is the set of, possibly fluctuating, abiotic environmental factors.
\end{itemize}
In other words, the quantities in $E_n$, $E_N$, $N_{1:P}$ and in $e$ define the biotic (intra- and inter-specific) and abiotic environments that the virtual strategy $x'$ is facing at time $t$, i.e., the quantities that are needed at time $t$ to determine the fitness of the strategy.
Similarly, I set
\begin{linenomath}
\begin{equation}
\label{eq:F}
F_M(n_{1:M},N_{1:P},x_{1:M},t):=F(E_N(x_{1:M},n_{1:M},N_{1:P},e(t)),N_{1:P},e(t)).
\end{equation}
\end{linenomath}
\end{subequations}

The point is now how the function sets $E_n$ and $E_N$ are defined over the distribution $\nu$.
Here is where I introduce the iterative procedure.

\subsection{The iterative procedure for building nonlinear fitnesses}
\label{ssec:proc}
Being constrained to smoothly operate on the distribution $\nu$ in \eqref{eq:nu}, only integrals over the whole strategy space are allowed.
Let me start with the intra-specific interactions described by the environmental quantities in $E$ and apply the integrals recursively as follows:
\begin{linenomath}
\begin{subequations}
\label{eq:Eq}
\begin{eqnarray}
\label{eq:E1}
E_n^{(1)}\hspace{-0.3mm}(x'\hspace{-0.7mm},x_{1:M}\hspace{-0.3mm},n_{1:M}\hspace{-0.3mm},e)
\hspace{-0.5mm} & := & \hspace{-1.2mm}
\int\hspace{-1.5mm} e_n^{(1)}(x'\hspace{-0.7mm},x,e)\hspace{0.5mm}\nu(x,x_{1:M}\hspace{-0.3mm},n_{1:M})\hspace{0.5mm}\mathrm{d}x\nonumber\\[-1.5mm]
 & & \hspace{-35.0mm}=
\sum_{j=1}^M e_n^{(1)}(x'\hspace{-0.7mm},x_j,e)\hspace{0.5mm}n_j,\\
\label{eq:E2}
E_n^{(2)}(x'\hspace{-0.7mm},x_{1:M}\hspace{-0.3mm},n_{1:M}\hspace{-0.3mm},N_{1:P},e)
\hspace{-0.5mm} & := & \hspace{-1.2mm}
\int\hspace{-1.5mm} e_n^{(2)}(x'\hspace{-0.7mm},x,\hspace{-0.5mm}E_n^{(1)}\hspace{-0.3mm}(x,x_{1:M}\hspace{-0.3mm},n_{1:M}\hspace{-0.3mm},e),N_{1:P},e)\nonumber\\[-0.5mm]
 & \times & 
\nu(x,x_{1:M}\hspace{-0.3mm},n_{1:M})\hspace{0.5mm}\mathrm{d}x\nonumber\\[-1.5mm]
 & & \hspace{-35.0mm}=
\sum_{j=1}^M e_n^{(2)}(x'\hspace{-0.7mm},x_j,\hspace{-0.5mm}E_n^{(1)}(x_j,x_{1:M}\hspace{-0.3mm},n_{1:M}\hspace{-0.3mm},e),N_{1:P},e)\hspace{0.5mm}n_j,\\[-2mm]
 & \hspace{0.5mm}\vdots & \nonumber\\
E_n^{(L)}(x'\hspace{-0.7mm},x_{1:M}\hspace{-0.3mm},n_{1:M}\hspace{-0.3mm},N_{1:P},e)
\hspace{-0.5mm} & := & \hspace{-1.2mm}
\int\hspace{-1.5mm} e_n^{(L)}(x'\hspace{-0.7mm},x,\hspace{-0.5mm}E_n^{(1:L-1)}(x,x_{1:M}\hspace{-0.3mm},n_{1:M}\hspace{-0.3mm},N_{1:P},e),N_{1:P},e)\nonumber\\[-0.5mm]
 & \times & 
\nu(x,x_{1:M}\hspace{-0.3mm},n_{1:M})\hspace{0.5mm}\mathrm{d}x\nonumber\\[-1.5mm]
 & & \hspace{-35.0mm}=
\sum_{j=1}^M e_n^{(L)}(x'\hspace{-0.7mm},x_j,\hspace{-0.5mm}E_n^{(1:L-1)}(x_j,x_{1:M}\hspace{-0.3mm},n_{1:M}\hspace{-0.3mm},N_{1:P},e),N_{1:P},e)\hspace{0.5mm}n_j,
\end{eqnarray}
\end{subequations}
\end{linenomath}
where $e_n^{(1)},\hspace{-0.6mm}...,e_n^{(L)}$ are suitable {\it interacting kernels}, i.e., vectors of functions (possibly affected by the abiotic environment $e$) describing the pairwise interactions of the virtual strategy $x'$ with the generic strategy $x$ spanning the strategy space
($E_n^{(1:L-1)}$ stands for $E_n^{(1)},\hspace{-0.6mm}...,\hspace{-0.5mm}E_n^{(L-1)}$).
And let me then collect all the environmental quantities $E_n^{(l)}$, $l\hspace{-0.5mm}=\hspace{-0.5mm}1,\hspace{-0.6mm}...,L$, in
\begin{linenomath}
\begin{equation}
\label{eq:E}
E_n(x'\hspace{-0.7mm},x_{1:M}\hspace{-0.3mm},n_{1:M}\hspace{-0.3mm},N_{1:P},e):=[E_n^{(1)}(x'\hspace{-0.7mm},x_{1:M}\hspace{-0.3mm},n_{1:M}\hspace{-0.3mm},e),\hspace{-0.6mm}...,\hspace{-0.3mm}E_n^{(L)}(x'\hspace{-0.7mm},x_{1:M}\hspace{-0.3mm},n_{1:M}\hspace{-0.3mm},N_{1:P},e)].
\end{equation}
\end{linenomath}
Note that $x$ is here treated, for simplicity, as a one-dimensional strategy, but this is not necessary.

The first integral step in \eqref{eq:E1} describes the interactions that are not affected by the concurrent intra-specific activities of the involved resident individual.
If such type of interactions contribute to the growth rate of the virtual strategy $x'$, then the first environmental component $E_n^{(1)}$ is a direct argument of function $g$ in \eqref{eq:G}.
The second integral step in \eqref{eq:E2} describes the interactions that are affected by the concurrent (intra-specific) activities of the involved $x_j$-strategist, and such dependence is accounted for by the $E_n^{(1)}$-dependence of $e_n^{(2)}$
($e_n^{(2)}$ might also depend on $N_{1:P}$, in case the concurrent $x_j$-activity extends to some of the other species, e.g. predation in Box~4).
Note that $E_n^{(1)}$ in the integral in \eqref{eq:E2} is evaluated for $x'=x$, and $x$ takes value $x_j$ in the resulting sum, so $E_n^{(1)}$ there describes the interactions of the $x_j$-strategist that are not affected by the concurrent (intra-specific) activities of the involved $x_k$-strategist, $k\neq j$, $j=1,\hspace{-0.6mm}...,\hspace{-0.3mm}M$.
If they were affected, then a further integral step would be required.
In this way, one can describe the interactions of the $x'$-strategist that are affected by the concurrent (intra-specific) activities of the involved resident individual up to a finite level of nesting.

A similar integral is used to define the set of functions $E_N$:
\begin{linenomath}
\begin{eqnarray}
\label{eq:EN}
E_N\hspace{-0.3mm}(x_{1:M}\hspace{-0.3mm},n_{1:M}\hspace{-0.3mm},\hspace{-0.2mm}N_{1:P},e) &:=&
\int\hspace{-1.5mm} e_N(x,E_n(x,x_{1:M}\hspace{-0.3mm},n_{1:M}\hspace{-0.3mm},\hspace{-0.2mm}N_{1:P},e),\hspace{-0.2mm}N_{1:P},e)\nonumber\\[-0.5mm]
 & \times & 
\nu(x,x_{1:M}\hspace{-0.3mm},n_{1:M})\hspace{0.5mm}\mathrm{d}x\nonumber\\[-0.5mm]
 & = & \!
\sum_{j=1}^M e_N(x_j,E_n(x_j,x_{1:M}\hspace{-0.3mm},n_{1:M}\hspace{-0.3mm},\hspace{-0.2mm}N_{1:P},e),\hspace{-0.2mm}N_{1:P},e)\hspace{0.5mm}n_j.
\end{eqnarray}
\end{linenomath}
Here only one step is sufficient.
When an individual of population $N_p$, $p=1,\hspace{-0.6mm}...,\hspace{-0.3mm}P$, encounters an $x_j$-strategist, the effect of their interaction can depend on the concomitant activities of the encountered individual.
The intra-specific ones are accounted for by the (possible) $E_n$-dependency of the interacting kernel $e_N$
(note the first argument $x$ of $E_n$ in the left-hand side of \eqref{eq:EN}, that takes value $x_j$ in sum at the right-hand side),
whereas other inter-specific activities can be described through the $N_{1:P}$-dependency of the growth rates in $g$ and $F$
(see Boxes~3 and~4 for two specific examples).

Of course, I do not claim that all smooth functions $g_M$ and $F_M$ in (\ref{eq:G},\hspace{0.2mm}b) can be obtained through the above procedure, but certainly all pairwise interactions can be described.

\subsection{The fitness' structural properties}
\label{ssec:prop}
In this section I present four structural properties that are satisfied by a $g$-function defined as in eq.~\eqref{eq:G} through the iterative procedure (\ref{eq:Eq}--\ref{eq:EN}).
\begin{enumerate}
\item[P1:]
$\begin{array}{rcl}
g_M(n_{1:M},N_{1:P},x_{1:M},x'\hspace{-0.5mm},t)\Big|_{n_1=0} & = & g_{M-1}(n_{2:M},N_{1:P},x_{2:M},x'\hspace{-0.5mm},t),
\end{array}$\\[1mm]
i.e., the fitness of the strategy $x'$ is not affected by the strategy $x_1$ of an absent population.
\vspace{2mm}
\item[P2:]
\raisebox{-6mm}{$\begin{array}{rcl}
g_M(n_{1:M},N_{1:P},x_{1:M},x'\hspace{-0.5mm},t)\Big|_{x_1=x_2=x}\\[2.5mm]
 & & \hspace{-50mm} =
g_M([\alpha(n_1\hspace{-0.8mm}+\hspace{-0.4mm}n_2),(1\hspace{-0.5mm}-\hspace{-0.3mm}\alpha)(n_1\hspace{-0.8mm}+\hspace{-0.4mm}n_2),
n_{3:M}],N_{1:P},[x,x,x_{3:M}],x'\hspace{-0.5mm},t)\\[1.5mm]
 & & \hspace{-50mm} =
g_{M-1}([n_1\hspace{-0.8mm}+\hspace{-0.4mm}n_2,n_{3:M}],N_{1:P},[x,x_{3:M}],x'\hspace{-0.5mm},t),
\end{array}$}\\[1mm]
for any $0\le\alpha\le 1$, i.e., any partitioning of the total density $n_1\hspace{-0.8mm}+\hspace{-0.4mm}n_2$ into two subpopulations with same strategy $x$ must result in the same per-capita growth rate for the strategy $x'$.
\vspace{2mm}
\item[P3:]
$\begin{array}{rcl}
g_M([n_2,n_1,n_{3:M}],N_{1:P},[x_2,x_1,x_{3:M}],x'\hspace{-0.5mm},t) & = & g_M(n_{1:M},N_{1:P},x_{1:M},x'\hspace{-0.5mm},t),
\end{array}$\\[1mm]
i.e., the order in which populations $1$ and $2$ are listed does not matter.
\vspace{2mm}
\item[P4:]
\raisebox{-6.5mm}{$\begin{array}{rcl}
\ds\left.\partial_{\hspace{-0.3mm}x_1^{d_1}\hspace{-1.8mm},\hspace{0.1mm}\ldots\hspace{0.1mm},\hspace{0.3mm}x_k^{d_k}}\hspace{0.5mm}
g_M(n_{1:M}\hspace{-0.3mm},\hspace{-0.5mm}N_{1:P},\hspace{-0.3mm}x_{1:M}\hspace{-0.3mm},\hspace{-0.3mm}x'\hspace{-0.5mm},\hspace{-0.3mm}t)\right|_{x_1=\cdots=x_M=x}\\[-1mm]
 & & \hspace{-60.0mm} = \ds
\sum_{i_1=1}^{d_1}\!\cdot\!\cdot\!\cdot\!\sum_{i_k=1}^{d_k}
\phi_{d_1\hspace{-0.2mm},\ldots,d_k,\hspace{0.3mm}i_1\hspace{-0.2mm},\ldots,i_k}\hspace{-0.3mm}(n,\hspace{-0.3mm}N_{1:P},\hspace{-0.3mm}x,\hspace{-0.3mm}x'\hspace{-0.5mm},t)\hspace{0.5mm}n_1^{i_1}\!\cdot\!\cdot\!\cdot n_k^{i_k},\\[1mm]
\end{array}$}\\[0.5mm]
for suitable functions $\phi$'s,
$d_1\hspace{-0.3mm},\hspace{-0.5mm}...,\hspace{-0.3mm}d_k\hspace{-0.5mm}\ge\hspace{-0.5mm}1$,\hspace{-0.3mm}
$1\hspace{-0.7mm}\le\hspace{-0.5mm}k\hspace{-0.5mm}\le\hspace{-0.7mm}M$,\hspace{-0.3mm}
and $n\hspace{-0.3mm}:=\hspace{-0.5mm}\sum_{j=1}^M\hspace{-0.5mm}n_j$.

This property is not obvious to be intuitively justified and is the main contribution of this work.
As anticipated in the Introduction, it generalizes the principle of mass-action, by assuming that $g_M$ describes pairwise interactions between the target $x'$-strategist and the resident strategies $x_{1:M}$ (or individuals of different species), which are, as well, involved in pairwise interactions.
As a consequence, when considering identical resident strategies, $x_1\!=\!x_2\!=\!\cdot\!\cdot\!\cdot\!=\!x_M\!=\!x$, the sensitivity of $g_M$ w.r.t.\hspace{-0.5mm} the resident strategies $x_j$ is proportional to the corresponding density $n_j$, with a proportionality coefficient that can be density-dependent only as a function of the total density $n$.
Moreover, due to nonlinear density dependencies in $g_M$, higher powers of $n_j$ may appear in further derivatives, up to the order $d_j$ of differentiation.
\end{enumerate}

Properties P1--4 can be generalized and combined to produce further relations among $g$-derivatives.
P1 and P2 obviously link $g_M$ to $g_{M-k}$ if $k$ resident strategies are absent and if $k+1$ resident strategies are equal, respectively.
In particular, by P2, we have
\begin{linenomath}
\begin{eqnarray}
\label{eq:P2p}
& &
g_M(n_{1:M},N_{1:P},x_{1:M},x'\hspace{-0.5mm},t)\Big|_{x_1=\cdots=x_M=x}=\nonumber\\[-0.5mm]
& &
g_{M-1}([n_1\hspace{-0.8mm}+\hspace{-0.4mm}n_2,n_{3:M}],N_{1:P},[x,x_{3:M}],x'\hspace{-0.5mm},t)\Big|_{x_3=\cdots=x_M=x}=\cdots=\nonumber\\[-1mm]
& &
g_{M\hspace{-0.2mm}-k}([n_1\hspace{-1.0mm}+\!\cdot\!\cdot\!\cdot\!+\hspace{-0.6mm}n_{k+1},n_{k+2:M}],N_{1:P},[x,x_{k+2:M}],x'\hspace{-0.5mm},t)\Big|_{x_{k+2}=\cdots=x_M=x}=\cdots=\nonumber\\
& &
g_1(n,N_{1:P},x,x'\hspace{-0.5mm},t),
\end{eqnarray}
\end{linenomath}
and derivatives w.r.t.\hspace{-0.5mm} the resident densities can be added, as the evaluation at $x_1=\cdots=x_M=x$ can be taken before the differentiation, thus obtaining:
\begin{linenomath}
\[
\partial_{n_1^{l_1}\hspace{-0.8mm},\hspace{0.3mm}\ldots,n_M^{l_M}}
g_M(n_{1:M},N_{1:P},x_{1:M},x'\hspace{-0.5mm},t)\Big|_{x_1=\cdots=x_M=x}\hspace{-1mm}=
\partial_{n^{l_1\hspace{-0.4mm}+\cdots+l_M}}
g_1(n,N_{1:P},x,x'\hspace{-0.5mm},t).
\]
\end{linenomath}
That is, $n_j$- and $n_k$-perturbations simply perturb the total density $n_j\hspace{-0.3mm}+n_k$ if the two populations have the same strategy $x$.

Property P3 obviously holds for any permutation of the elements in $n_{1:M}$ and in $x_{1:M}$ in the left-hand side.
And thanks to P3, also P4 can be generalized to derivatives taken to any set of $k$ resident strategies, not necessarily the first $k$.
Moreover, derivatives w.r.t.\hspace{-0.5mm} the densities $N_{1:P}$ and w.r.t.\hspace{-0.5mm} the virtual strategy $x'$ can be added to all properties.

Three further remarks are due for the functions $\phi$'s.
First note that they do not depend on the number $M$ of considered residents, but they are rather associated to the considered focus species.
E.g., the same functions $\phi_{d,i}$'s, characterized by a single sum index ($1$-index-$\phi$'s in the following), appear when applying P4 to a pure-derivative of $g_M$ and of $g_1$, i.e.,
\begin{eqnarray*}
\partial_{x_k^d}\hspace{0.3mm}
g_M(n_{1:M},N_{1:P},x_{1:M},x'\hspace{-0.5mm},t)\Big|_{x_1=\cdots=x_M=x}\hspace{-1mm} & = &
\sum_{i=1}^d\phi_{d,i}(n,\hspace{-0.2mm}N_{1:P},x,x'\hspace{-0.5mm},t)\hspace{0.5mm}n_k^i,\\
\partial_{x}\hspace{0.3mm}
g_1(n,\hspace{-0.2mm}N_{1:P},x,x'\hspace{-0.5mm},t)  & = & \ds
\sum_{i=1}^d\phi_{d,i}(n,\hspace{-0.2mm}N_{1:P},x,x'\hspace{-0.5mm},t)\hspace{0.5mm}n^i,
\end{eqnarray*}
$1\hspace{-0.7mm}\le\hspace{-0.5mm}k\hspace{-0.5mm}\le\hspace{-0.7mm}M$.
Analogously, the same functions $\phi_{d_1\hspace{-0.2mm},\ldots,d_k,\hspace{0.3mm}i_1\hspace{-0.2mm},\ldots,i_k}\hspace{-0.5mm}$'s ($k$-index-$\phi$'s) appear when perturbing any set of $k$ different resident strategies.

Second, by property P3, we obviously have
\begin{linenomath}
\begin{equation}
\label{eq:P34}
\phi_{d_2,d_1\hspace{-0.2mm},d_3,\ldots,d_k,\hspace{0.3mm}i_2,i_1\hspace{-0.2mm},i_3,\ldots,i_k}=
\phi_{d_1\hspace{-0.2mm},d_2,d_3,\ldots,d_k,\hspace{0.3mm}i_1\hspace{-0.2mm},i_2,i_3,\ldots,i_k},
\end{equation}
\end{linenomath}
that easily generalizes to any permutation of the indexes $d_1\hspace{-0.2mm},\hspace{-0.6mm}...,\hspace{-0.2mm}d_k$ and $i_1\hspace{-0.2mm},\hspace{-0.6mm}...,\hspace{-0.2mm}i_k$ in the left-hand side.

Third, there are also links between the $k$- and $k'$-index-$\phi$'s, $k'<k$, of the same order
(same sums $d:=d_1\hspace{-0.8mm}+\!\cdot\!\cdot\!\cdot\!+\hspace{-0.3mm}d_k\ge k$ and $d_1\hspace{-0.8mm}+\!\cdot\!\cdot\!\cdot\!+\hspace{-0.3mm}d_{k'}$).
They come from properties P2 and P4, by $x$-differentiating \eqref{eq:P2p} $d$ times, applying P4, and then collecting the same powers of the resident densities at both sides of an equal sign.
E.g., for $d=2$ and considering the left- and right-most sides of \eqref{eq:P2p}, we get
\begin{linenomath}
\begin{eqnarray*}
& & \hspace{-2mm}
\sum_{i=1}^{M}\partial_{x_i^2}g_M(n_{1:M},N_{1:P},x_{1:M},x'\hspace{-0.5mm},t)\Big|_{x_1=\cdots=x_M=x}+\\
& &
2\!\sum_{i=1}^{M-1}\!\!\sum_{j=i+1}^{M}\!\!\partial_{x_i,x_j}g_M(n_{1:M},N_{1:P},x_{1:M},x'\hspace{-0.5mm},t)\Big|_{x_1=\cdots=x_M=x}\hspace{-1.0mm}=
\partial_{x^2}g_1(n,N_{1:P},x,x'\hspace{-0.5mm},t),
\end{eqnarray*}
\end{linenomath}
which becomes by P4
\begin{linenomath}
\begin{eqnarray*}
& & \hspace{-2mm}
\sum_{i=1}^{M}\!\Big(\hspace{-0.3mm}\phi_{2,1}\hspace{-0.3mm}(n,N_{1:P},x,x'\hspace{-0.5mm},t)\hspace{0.2mm}n_i+
\phi_{2,2}(n,N_{1:P},x,x'\hspace{-0.5mm},t)\hspace{0.1mm}n_i^2\Big)+\\
& &
2\!\sum_{i=1}^{M-1}\!\!\sum_{j=i+1}^{M}\!\!\phi_{1,1,1,1}\hspace{-0.3mm}(n,N_{1:P},x,x'\hspace{-0.5mm},t)\hspace{0.2mm}n_i\hspace{0.2mm}n_j=\\
& &
\phi_{2,1}(n,N_{1:P},x,x'\hspace{-0.5mm},t)\sum_{i=1}^{M}n_i + 
\phi_{2,2}(n,N_{1:P},x,x'\hspace{-0.5mm},t)\Big(\sum_{i=1}^{M}n_i^2+2\!\sum_{i=1}^{M-1}\!\!\sum_{j=i+1}^{M}n_i\hspace{0.2mm}n_j\Big).
\end{eqnarray*}
\end{linenomath}
Then, the $n_i$- and $n_i^2$-terms cancel out in the above equation, whereas balancing the mixed $n_i\hspace{0.2mm}n_j$-terms yields
\begin{subequations}
\label{eq:phir}
\begin{equation}
\label{eq:phi1111}
\phi_{1,1,1,1}=\phi_{2,2}.
\end{equation}
For $d=3$, we get
\begin{linenomath}
\begin{eqnarray*}
& & \hspace{-3mm}
\sum_{i=1}^{M}\partial_{x_i^3}g_M(n_{1:M},N_{1:P},x_{1:M},x'\hspace{-0.5mm},t)\Big|_{x_1=\cdots=x_M=x}+\\
& &
3\!\sum_{i=1}^{M-1}\!\!\sum_{j=i+1}^{M}\!\!
\partial_{x_i^2,x_j}g_M(n_{1:M},N_{1:P},x_{1:M},x'\hspace{-0.5mm},t)\Big|_{x_1=\cdots=x_M=x}+\\
& &
3\!\sum_{i=1}^{M-1}\!\!\sum_{j=i+1}^{M}\!\!
\partial_{x_i,x_j^2}g_M(n_{1:M},N_{1:P},x_{1:M},x'\hspace{-0.5mm},t)\Big|_{x_1=\cdots=x_M=x}+\\
& &
6\!\sum_{i=1}^{M-2}\!\sum_{\hspace{-0.5mm}j=i+1}^{M-1}\hspace{-3.5mm}\sum_{\hspace{3mm}k=j+1}^{M}\hspace{-3mm}
\partial_{x_i,x_j,x_k}g_M(n_{1:M},N_{1:P},x_{1:M},x'\hspace{-0.5mm},t)\Big|_{x_1=\cdots=x_M=x}\hspace{-1.0mm}=\\
& & \hspace{-3mm}
\partial_{x^3}g_1(n,N_{1:P},x,x'\hspace{-0.5mm},t),
\end{eqnarray*}
\end{linenomath}
which becomes by P4
\begin{linenomath}
\begin{eqnarray*}
& & \hspace{-3mm}
\sum_{i=1}^{M}\!\Big(\hspace{-0.3mm}\phi_{3,1}\hspace{-0.3mm}(n,N_{1:P},x,x'\hspace{-0.5mm},t)\hspace{0.2mm}n_i+
\phi_{3,2}(n,N_{1:P},x,x'\hspace{-0.5mm},t)\hspace{0.1mm}n_i^2+
\phi_{3,3}(n,N_{1:P},x,x'\hspace{-0.5mm},t)\hspace{0.1mm}n_i^2\Big)+\\
& &
3\!\sum_{i=1}^{M-1}\!\!\sum_{j=i+1}^{M}\!\Big(\hspace{-0.3mm}\phi_{2,1,1,1}\hspace{-0.3mm}(n,N_{1:P},x,x'\hspace{-0.5mm},t)\hspace{0.2mm}n_i\hspace{0.2mm}n_j+\phi_{2,1,2,1}\hspace{-0.3mm}(n,N_{1:P},x,x'\hspace{-0.5mm},t)\hspace{0.2mm}n_i^2\hspace{0.2mm}n_j\Big)\\
& &
3\!\sum_{i=1}^{M-1}\!\!\sum_{j=i+1}^{M}\!\Big(\hspace{-0.3mm}\phi_{1,2,1,1}\hspace{-0.3mm}(n,N_{1:P},x,x'\hspace{-0.5mm},t)\hspace{0.2mm}n_i\hspace{0.2mm}n_j+\phi_{1,2,1,2}\hspace{-0.3mm}(n,N_{1:P},x,x'\hspace{-0.5mm},t)\hspace{0.2mm}n_i\hspace{0.2mm}n_j^2\Big)\\
& &
6\!\sum_{i=1}^{M-2}\!\sum_{\hspace{-0.5mm}j=i+1}^{M-1}\hspace{-3.5mm}\sum_{\hspace{3mm}k=j+1}^{M}\!\!
\phi_{1,1,1,1,1,1}\hspace{-0.3mm}(n,N_{1:P},x,x'\hspace{-0.5mm},t)\hspace{0.2mm}n_i\hspace{0.2mm}n_j\hspace{0.2mm}n_k=\\
& & \hspace{-3mm}
\phi_{3,1}(n,N_{1:P},x,x'\hspace{-0.5mm},t)\sum_{i=1}^{M}n_i + 
\phi_{3,2}(n,N_{1:P},x,x'\hspace{-0.5mm},t)\Big(\sum_{i=1}^{M}n_i^2+2\!\sum_{i=1}^{M-1}\!\!\sum_{j=i+1}^{M}n_i\hspace{0.2mm}n_j\Big) +\\
& & \hspace{-3mm}
\phi_{3,3}(n,N_{1:P},x,x'\hspace{-0.5mm},t)\Big(\sum_{i=1}^{M}n_i^3+
3\!\sum_{i=1}^{M-1}\!\!\sum_{j=i+1}^{M}\hspace{-2mm}\big(n_i^2\hspace{0.2mm}n_j\hspace{-0.5mm}+\hspace{-0.2mm}n_i\hspace{0.2mm}n_j^2\big)+
6\!\sum_{i=1}^{M-2}\!\sum_{\hspace{-0.5mm}j=i+1}^{M-1}\hspace{-3.5mm}\sum_{\hspace{3mm}k=j+1}^{M}\hspace{-3mm}n_i\hspace{0.2mm}n_j\hspace{0.2mm}n_k\Big).
\end{eqnarray*}
\end{linenomath}
Similarly, the $n_i$-, $n_i^2$-, and $n_i^3$-terms cancel out, whereas balancing the mixed $n_i\hspace{0.2mm}n_j$-, $n_i^2\hspace{0.2mm}n_j$-, $n_i\hspace{0.2mm}n_j^2$-, and $n_i\hspace{0.2mm}n_j\hspace{0.2mm}n_k$-terms yields
\begin{linenomath}
\begin{eqnarray}
\phi_{2,1,1,1}\overset{P3}{=}\phi_{1,2,1,1} &=& \onethird\phi_{3,2},\\[-1mm]
\phi_{2,1,2,1}\overset{P3}{=}\phi_{1,2,1,2} &=& \phi_{3,3},\\
\phi_{1,1,1,1,1,1} &=& \phi_{3,3}
\end{eqnarray}
\end{linenomath}
\end{subequations}
(basically, $\phi_{d_1\hspace{-0.2mm},\ldots,d_k,\hspace{0.3mm}d_1\hspace{-0.2mm},\ldots,d_k}\hspace{-0.3mm}=\phi_{d,d}$ holds at any order, whereas suitable combinations link the $\phi_{d_1\hspace{-0.2mm},\ldots,d_k,\hspace{0.3mm}i_1\hspace{-0.2mm},\ldots,i_k}$'s with some $i_j\hspace{-0.5mm}<\hspace{-0.3mm}d_j$, $1\hspace{-0.5mm}\ge\hspace{-0.3mm}j\hspace{-0.5mm}\ge\hspace{-0.3mm}k$, with $\phi_{d,i}$, $i:=i_1\hspace{-0.8mm}+\!\cdot\!\cdot\!\cdot\!+\hspace{-0.3mm}i_k$).

In this paper, I am not focusing on the relations between (same-order) $k$- and $k'$-index-$\phi$'s, $k'<k$.
The important message here is that the $g$-derivatives w.r.t.\hspace{-0.5mm} the resident strategies can be organized as in P4, i.e., as a sum of terms each composed by a coefficient $\phi$, function of the total density $n$ of the species, times powers of the densities of the perturbed strategies, up to the order of differentiation.
In Sect.~\ref{ssec:how} I will show how all functions $\phi$'s can be directly identified, without making use of their relations.
Those relations do exist---they can all be derived as exemplified above---and will be satisfied by construction by the identified $\phi$'s.

Finally, analogous properties obviously hold for the population growth rates of the other species in $F_M$:
\begin{enumerate}
\item[P1]
$\begin{array}{rcl}
F_M(n_{1:M},N_{1:P},x_{1:M},t)\Big|_{n_1=0} & = & F_{M-1}(n_{2:M},N_{1:P},x_{2:M},t),
\end{array}$
\vspace{1mm}
\item[P2]
\raisebox{-6.1mm}{$\begin{array}{rcl}
F_M(n_{1:M},N_{1:P},x_{1:M},t)\Big|_{x_1=x_2=x}\\[2.5mm]
 & & \hspace{-50mm} =
F_M([\alpha(n_1\hspace{-0.8mm}+\hspace{-0.4mm}n_2),(1\hspace{-0.5mm}-\hspace{-0.3mm}\alpha)(n_1\hspace{-0.8mm}+\hspace{-0.4mm}n_2),n_{3:M}],N_{1:P},[x,x,x_{3:M}],t)\\[1.5	mm]
 & & \hspace{-50mm} =
F_{\hspace{-0.3mm}M-1\hspace{-0.3mm},\hspace{0.3mm}p}([n_1\hspace{-0.8mm}+\hspace{-0.4mm}n_2,n_{3:M}],N_{1:P},[x,x_{3:M}],t),
\end{array}$}
\vspace{1mm}
\item[P3]
$\begin{array}{rcl}
F_M([n_2,n_1,n_{3:M}],N_{1:P},[x_2,x_1,x_{3:M}],t) & = & F_M(n_{1:M},N_{1:P},x_{1:M},t),
\end{array}$
\vspace{1mm}
\item[P4]
\raisebox{-6.5mm}{$\begin{array}{rcl}
\ds\left.\partial_{\hspace{-0.3mm}x_1^{d_1}\hspace{-1.8mm},\hspace{0.1mm}\ldots\hspace{0.1mm},\hspace{0.3mm}x_k^{d_k}}\hspace{0.3mm}
F_M(n_{1:M}\hspace{-0.3mm},\hspace{-0.5mm}N_{1:P},\hspace{-0.3mm}x_{1:M}\hspace{-0.3mm},\hspace{-0.3mm}t)\right|_{x_1=\cdots=x_M=x}\\[-1mm]
 & & \hspace{-60.0mm} = \ds
\sum_{i_1=1}^{d_1}\!\cdot\!\cdot\!\cdot\!\sum_{i_k=1}^{d_k}
\psi_{d_1\hspace{-0.2mm},\ldots,d_k,\hspace{0.3mm}i_1\hspace{-0.2mm},\ldots,i_k}\hspace{-0.3mm}(n,\hspace{-0.3mm}N_{1:P},\hspace{-0.3mm}x,\hspace{-0.3mm}t)\hspace{0.5mm}n_1^{i_1}\!\cdot\!\cdot\!\cdot n_k^{i_k},
\end{array}$}
\end{enumerate}
where $\psi$'s are suitable vectors of functions.
And analogous generalization and combinations go through, including the relations between (same-order) $k$- and $k'$-index-$\psi$'s, $k'<k$, such as
\begin{linenomath}
\begin{subequations}
\label{eq:psir}
\begin{eqnarray}
\label{eq:psi1111}
\psi_{1,1,1,1} &=& \psi_{2,2},\\[-1mm]
\psi_{2,1,1,1}\overset{P3}{=}\psi_{1,2,1,1} &=& \onethird\psi_{3,2},\\[-1mm]
\psi_{2,1,2,1}\overset{P3}{=}\psi_{1,2,1,2} &=& \psi_{3,3},\\
\psi_{1,1,1,1,1,1} &=& \psi_{3,3}.
\end{eqnarray}
\end{subequations}
\end{linenomath}
However, in the following, I mainly discuss the properties of the $g$-function associated to the focus species.

\subsection{How the procedure secures the properties}
\label{ssec:how}
I now show that the definition of $g_M$ (\ref{eq:G}) and the iterative procedure (\ref{eq:Eq}--\ref{eq:EN}) ensure properties P1--4.
I do not show the converse---a $g$-function satisfying P1--4 can be defined as in (\ref{eq:G},~\ref{eq:Eq}--\ref{eq:EN}) with suitable interacting kernels---though I believe it is true.

Properties P1--4 follow directly by the integral procedure (\ref{eq:Eq}--\ref{eq:EN}).
In fact, integrating functionals of the distribution $\nu$ over all possible resident strategies implies that:
the strategy $x_i$ of an absent population ($n_i=0$) has no effect (P1);
the densities of equal strategies can be summed up ($n_1\hspace{0.1mm}\delta(x\hspace{-0.2mm}-\hspace{-0.2mm}x_1)+n_2\hspace{0.35mm}\delta(x\hspace{-0.2mm}-\hspace{-0.2mm}x_2)=(n_1\hspace{-0.4mm}+\hspace{-0.2mm}n_2)\hspace{0.35mm}\delta(x\hspace{-0.2mm}-\hspace{-0.2mm}x_1)$ if $x_1\hspace{-0.4mm}=\hspace{-0.2mm}x_2$) (P2);
the order of the peaks in the sum in \eqref{eq:nu} is irrelevant (P3).

Property P4 can be easily verified by means of symbolic computations on specific examples
(see the {\it Mathematica} script provided as Supplementary Material for the examples described in Boxes~3 and~4).
Specifically, given P4 as granted, the functions $\phi_{d_1\hspace{-0.2mm},\ldots,d_k,\hspace{0.3mm}i_1\hspace{-0.2mm},\ldots,i_k}$'s can be uniquely identified as follows.
For any given set set $d_1,\hspace{-0.6mm}...,d_k$, $k\ge1$, take the $\big(\partial\,{\raisebox{-0mm}{$\ss\hspace{-0.3mm}x_1^{d_1}\hspace{-1.8mm},\hspace{0.1mm}\ldots\hspace{0.1mm},\hspace{0.3mm}x_k^{d_k}$}}\big)$-derivative of $g_{k+1}$ (i.e., $M\hspace{-0.5mm}=\hspace{-0.3mm}k\hspace{-0.2mm}+\hspace{-0.5mm}1$) and replace, whenever possible, $n_1\hspace{-0.7mm}+\!\cdot\!\cdot\!\cdot\!+\hspace{-0.3mm}n_{k+1}$ with $n$.
This can be systematically done by noting that the density $n_{k+1}$ appears in the resulting derivative only in sum with all other resident densities $n_1,\hspace{-0.6mm}...,n_k$, since no differentiation is taken w.r.t.\hspace{-0.5mm} $x_{k+1}$.
Then, by replacing $n_{k+1}$ with $n-(n_1\hspace{-0.7mm}+\!\cdot\!\cdot\!\cdot\!+\hspace{-0.3mm}n_{k})$, the sums where $n_{k+1}$ was appearing before the substitution simplifies to $n$, whereas all the remaining occurrences of $n_1,\hspace{-0.6mm}...,n_k$ form the monomials $n_1^{i_1}\!\cdot\!\cdot\!\cdot n_k^{i_k}$ in P4.
Now, for each set $i_1,\hspace{-0.6mm}...,i_k$, $1\hspace{-0.5mm}\le\hspace{-0.2mm}i_j\hspace{-0.6mm}\le\hspace{-0.6mm}d_j$, $j\hspace{-0.5mm}=\hspace{-0.5mm}1\hspace{-0.1mm},\hspace{-0.6mm}...,k$, function $\phi_{d_1\hspace{-0.2mm},\ldots,d_k,\hspace{0.3mm}i_1\hspace{-0.2mm},\ldots,i_k}$ can then be obtained collecting the coefficient of the corresponding monomial.
Similarly for function $\psi_{d_1\hspace{-0.2mm},\ldots,d_k,\hspace{0.3mm}i_1\hspace{-0.2mm},\ldots,i_k}$.

Below I show, as a general case, that
\begin{linenomath}
\begin{equation}
\label{eq:phi11}
\left.\partial_{x_1^{\phantom{d}}}\hspace{0.2mm}
g_M(n_{1:M}\hspace{-0.3mm},\hspace{-0.5mm}N_{1:P},\hspace{-0.3mm}x_{1:M}\hspace{-0.3mm},\hspace{-0.3mm}x'\hspace{-0.5mm},\hspace{-0.3mm}t)\right|_{x_1=\cdots=x_M=x}=\hspace{0.5mm}
\phi_{1,1}\hspace{-0.3mm}(n,\hspace{-0.3mm}N_{1:P},\hspace{-0.3mm}x,\hspace{-0.3mm}x'\hspace{-0.5mm},t)\hspace{0.5mm}n_1,
\end{equation}
\end{linenomath}
i.e., property P4 for $k=1$ and $d_1=1$.

From definition \eqref{eq:G} and the chain rule, it results
\begin{linenomath}
\begin{eqnarray}
\label{eq:phi11_1}
& & \hspace{-6.0mm}
\Frac{\raisebox{-0.8mm}{$1$}}{\raisebox{1.0mm}{$\hspace{0.5mm}n_1\hspace{-0.5mm}$}}\hspace{-0.5mm}
\left.\partial_{x_1^{\phantom{d}}}
g_M(n_{1:M}\hspace{-0.3mm},\hspace{-0.5mm}N_{1:P},\hspace{-0.3mm}x_{1:M}\hspace{-0.3mm},\hspace{-0.3mm}x'\hspace{-0.5mm},\hspace{-0.3mm}t)\right|_{x_1=\cdots=x_M=x}=\nonumber\\
& & \hspace{-6.0mm}
\Frac{\raisebox{-0.8mm}{$1$}}{\raisebox{1.0mm}{$\hspace{0.5mm}n_1\hspace{-0.5mm}$}}\hspace{-0.5mm}
\left.\partial_{x_1^{\phantom{d}}}
g(x'\hspace{-0.5mm},E_n(x'\hspace{-0.5mm},x_{1:M},n_{1:M},\hspace{-0.4mm}N_{1:P},e),E_N(x_{1:M},n_{1:M},\hspace{-0.4mm}N_{1:P},e),\hspace{-0.2mm}N_{1:P},e\hspace{-0.5mm})\right|_{x_1=\cdots=x_M=x}=\nonumber\\
& & \hspace{-6.0mm}
\hspace{0.0mm}\left(\hspace{-0.9mm}\left.\partial_{E_n}\hspace{0.5mm}
g(x'\hspace{-0.8mm},\hspace{-0.3mm}E_n,E_N,\hspace{-0.3mm}N_{1:P},\hspace{-0.3mm}e\hspace{-0.3mm})\Big|
\raisebox{0mm}{\hspace{-1.5mm}
\begin{minipage}[t]{40mm}
$\ss E_n=E_n(x'\hspace{-0.7mm},\hspace{0.3mm}x_{1:M}\hspace{-0.4mm},\hspace{0.3mm}n_{1:M}\hspace{-0.4mm},\hspace{0.1mm}N_{1:P}\hspace{-0.3mm},\hspace{0.1mm}e)$\\[-2mm]
$\ss E_N=E_N(x_{1:M}\hspace{-0.4mm},\hspace{0.3mm}n_{1:M}\hspace{-0.4mm},\hspace{0.1mm}N_{1:P}\hspace{-0.3mm},\hspace{0.1mm}e)$
\end{minipage}}
\hspace{-1.0mm}
\Frac{\raisebox{-0.8mm}{$1$}}{\raisebox{1.0mm}{$\hspace{0.5mm}n_1\hspace{-0.5mm}$}}
\hspace{0.2mm}\partial_{x_1^{\phantom{d}}}\hspace{-0.5mm}
E_n(x'\hspace{-0.8mm},\hspace{-0.3mm}x_{1:M}\hspace{-0.3mm},\hspace{-0.3mm}n_{1:M}\hspace{-0.3mm},\hspace{-0.4mm}N_{1:P},\hspace{-0.3mm}e\hspace{-0.3mm})\hspace{-0.9mm}\right)\hspace{-0.7mm}\right|_{x_1=\cdots=x_M=x}+\nonumber\\
& & \hspace{-6.0mm}
\left(\hspace{-0.9mm}\left.\partial_{E_N}
g(x'\hspace{-0.8mm},\hspace{-0.3mm}E_n,E_N,\hspace{-0.3mm}N_{1:P},\hspace{-0.3mm}e\hspace{-0.3mm})\Big|
\raisebox{0mm}{\hspace{-1.5mm}
\begin{minipage}[t]{40mm}
$\ss E_n=E_n(x'\hspace{-0.7mm},\hspace{0.3mm}x_{1:M}\hspace{-0.4mm},\hspace{0.3mm}n_{1:M}\hspace{-0.4mm},\hspace{0.1mm}N_{1:P}\hspace{-0.3mm},\hspace{0.1mm}e)$\\[-2mm]
$\ss E_N=E_N(x_{1:M}\hspace{-0.4mm},\hspace{0.3mm}n_{1:M}\hspace{-0.4mm},\hspace{0.1mm}N_{1:P}\hspace{-0.3mm},\hspace{0.1mm}e)$
\end{minipage}}
\hspace{-1.0mm}
\Frac{\raisebox{-0.8mm}{$1$}}{\raisebox{1.0mm}{$\hspace{0.5mm}n_1\hspace{-0.5mm}$}}
\hspace{0.2mm}\partial_{x_1^{\phantom{d}}}\hspace{-0.5mm}
E_N(x_{1:M}\hspace{-0.3mm},\hspace{-0.3mm}n_{1:M}\hspace{-0.3mm},\hspace{-0.4mm}N_{1:P},\hspace{-0.3mm}e\hspace{-0.3mm})\hspace{-0.9mm}\right)\hspace{-0.7mm}\right|_{x_1=\cdots=x_M=x},\nonumber\\
\end{eqnarray}
\end{linenomath}
where
\begin{subequations}
\label{eq:dE}
\begin{linenomath}
\begin{eqnarray}
\label{eq:dE1}
\Frac{\raisebox{-0.8mm}{$1$}}{\raisebox{1.0mm}{$\hspace{0.5mm}n_1\hspace{-0.5mm}$}}\hspace{-0.3mm}
\left.\partial_{x_1^{\phantom{d}}}\hspace{-0.5mm}
E_n^{(1)}(x'\hspace{-0.5mm},x_{1:M},n_{1:M},e)\right|_{x_1=\cdots=x_M=x}
\hspace{-1.mm} &=& \hspace{-0.6mm}
\left.\partial_{x_1^{\phantom{d}}}\hspace{-0.2mm}
e_n^{(1)}(x'\hspace{-0.5mm},x_1,e)\right|_{x_1=x}
\end{eqnarray}
\end{linenomath}
and, for $L>1$,
\begin{linenomath}
\begin{eqnarray}
\label{eq:dEq}
& & \hspace{-12.0mm}
\Frac{\raisebox{-0.8mm}{$1$}}{\raisebox{1.0mm}{$n_1\hspace{-1.0mm}$}}
\hspace{-0.5mm}\left.\partial_{x_1^{\phantom{d}}}\hspace{-0.2mm}
E_n^{(L)}(x'\hspace{-0.5mm},x_{1:M},n_{1:M},\hspace{-0.2mm}N_{1:P},e)\right|_{x_1=\cdots=x_M=x}\hspace{-0.5mm}=
\Frac{\raisebox{-0.8mm}{$1$}}{\raisebox{1.0mm}{$n_1\hspace{-1.0mm}$}}
\bigg[
\partial_{x_1}e_n^{(L)}(x'\hspace{-0.5mm},x_1,E_n^{(1)},\hspace{-0.7mm}...,\hspace{-0.3mm}E_n^{(L-\hspace{-0.2mm}1)},\hspace{-0.2mm}N_{1:P},e)\hspace{0.2mm}n_1\nonumber\\
& & \hspace{-6.0mm}+\,
\partial_{E_n^{(1)}}\hspace{0.0mm}e_n^{(L)}(x'\hspace{-0.5mm},x,E_n^{(1)},\hspace{-0.7mm}...,\hspace{-0.3mm}E_n^{(L-\hspace{-0.2mm}1)},\hspace{-0.2mm}N_{1:P},e)\nonumber\\[-1.5mm]
& & \hspace{0.0mm}
\hspace{-0.5mm}\Big(\hspace{-0.5mm}
\partial_{x_1}E_n^{(1)}(x_1,\hspace{-0.5mm}[x,\hspace{-0.7mm}...,\hspace{-0.3mm}x],n_{1:M},\hspace{-0.3mm}e)\hspace{0.2mm}n_1
\hspace{-0.5mm}+\hspace{-0.5mm}\makebox[5.5mm][l]{$\sum\limits_{j=1}^M$}
\partial_{x_1}E_n^{(1)}(x\hspace{-0.1mm},\hspace{-0.2mm}x_{1:M}\hspace{-0.2mm},n_{1:M},\hspace{-0.2mm}e)\hspace{0.5mm}n_j\hspace{-0.5mm}\Big)\nonumber\\[-4mm]
& & \hspace{-5.0mm}
\vdots\nonumber\\[-1mm]
& & \hspace{-6.0mm}+\,
\partial_{E_n^{(L-\hspace{-0.2mm}1)}}\hspace{0.0mm}e_n^{(L)}(x'\hspace{-0.5mm},x,E_n^{(1)},\hspace{-0.7mm}...,\hspace{-0.3mm}E_n^{(L-\hspace{-0.2mm}1)},\hspace{-0.2mm}N_{1:P},e)
\hspace{-0.0mm}\Big(\hspace{-0.7mm}
\partial_{x_1}\hspace{-0.2mm}E_n^{(L-\hspace{-0.2mm}1)}\hspace{-0.3mm}(x_1\hspace{-0.2mm},\hspace{-0.7mm}[x,\hspace{-0.7mm}...,\hspace{-0.3mm}x],n_{1:M},\hspace{-0.2mm}N_{1:P},\hspace{-0.5mm}e)\hspace{0.2mm}n_1\nonumber\\
& & + \makebox[5.5mm][l]{$\sum\limits_{j=1}^M$}
\partial_{x_1}\hspace{-0.3mm}E_n^{(L-\hspace{-0.2mm}1)}\hspace{-0.2mm}(x\hspace{-0.1mm},\hspace{-0.2mm}x_{1:M}\hspace{-0.2mm},n_{1:M},\hspace{-0.2mm}N_{1:P},e)\hspace{0.2mm}n_j
\hspace{-0.6mm}\Big)
\hspace{-0.8mm}\bigg]\hspace{-0.5mm}\raisebox{-1.0mm}{\Bigg|}\raisebox{0mm}{\hspace{-1.0mm}
\begin{minipage}[t]{57mm}
$\ss\hspace{-0.5mm} E_n^{(1)}=E_n^{(1)}(\hspace{-0.1mm}x'\hspace{-0.9mm},\hspace{0.3mm}x_{1:M}\hspace{-0.4mm},\hspace{0.2mm}n_{1:M}\hspace{-0.4mm},\hspace{0.1mm}e)$\\[-1.2mm]
$\ss\hspace*{-0.5mm} E_n^{(l)}\hspace{0.3mm}=E_n^{(l)}(\hspace{-0.1mm}x'\hspace{-0.9mm},\hspace{0.3mm}x_{1:M}\hspace{-0.4mm},\hspace{0.2mm}n_{1:M}\hspace{-0.4mm},\hspace{0.1mm}N_{1:P},\hspace{0.1mm}e),\,
l=2,\hspace{-0.7mm}...,\hspace{0.3mm}L-\hspace{-0.2mm}1$\\[-2mm]
$\ss x_1=\cdots=\hspace{0.3mm}x_M=\hspace{0.3mm}x$
\end{minipage}}\nonumber\\
& & \hspace{-12.0mm}=
\bigg[
\partial_{x_1}e_n^{(L)}(x'\hspace{-0.5mm},x_1,E_n^{(1)},\hspace{-0.7mm}...,\hspace{-0.3mm}E_n^{(L-\hspace{-0.2mm}1)},\hspace{-0.2mm}N_{1:P},e)\hspace{0.5mm}\nonumber\\
& & \hspace{-6.0mm}+\hspace{0.5mm}n\hspace{0.5mm}
\partial_{E_n^{(1)}}\hspace{0.0mm}e_n^{(L)}(x'\hspace{-0.5mm},x,E_n^{(1)},\hspace{-0.7mm}...,\hspace{-0.3mm}E_n^{(L-\hspace{-0.2mm}1)},\hspace{-0.2mm}N_{1:P},e)\nonumber\\[0mm]
& & \hspace{0.0mm}
\hspace{-0.5mm}\Big(\hspace{-0.5mm}
\partial_{x_1}e_1(x_1,x,e)
\hspace{-0.5mm}+\hspace{-0.5mm}
\Frac{\raisebox{-0.8mm}{$1$}}{\raisebox{1.0mm}{$n_1\hspace{-1.0mm}$}}
\hspace{0.5mm}
\partial_{x_1}E_n^{(1)}(x\hspace{-0.1mm},\hspace{-0.2mm}x_{1:M}\hspace{-0.2mm},n_{1:M},\hspace{-0.2mm}e)
\hspace{-0.5mm}\Big)\nonumber\\[-4mm]
& & \hspace{-5.0mm}
\vdots\nonumber\\[-1mm]
& & \hspace{-6.0mm}+\hspace{0.5mm}n\hspace{0.5mm}
\partial_{E_n^{(L-\hspace{-0.2mm}1)}}\hspace{0.0mm}e_n^{(L)}(x'\hspace{-0.5mm},x,E_n^{(1)},\hspace{-0.7mm}...,\hspace{-0.3mm}E_n^{(L-\hspace{-0.2mm}1)},\hspace{-0.2mm}N_{1:P},e)
\hspace{-0.5mm}\Big(\hspace{-0.5mm}
\partial_{x_1}e_n^{(L-\hspace{-0.2mm}1)}\hspace{-0.2mm}(x_1\hspace{-0.2mm},\hspace{-0.2mm}x,\hspace{-0.5mm}E_n^{(1)},\hspace{-0.7mm}...,\hspace{-0.5mm}E_n^{(L-\hspace{-0.2mm}2)},\hspace{-0.2mm}N_{1:P},e)
\hspace{0.0mm}\nonumber\\
& & \hspace{-0.0mm}+\hspace{0.5mm}
\Frac{\raisebox{-0.8mm}{$1$}}{\raisebox{1.0mm}{$n_1\hspace{-1.0mm}$}}
\hspace{0.5mm}
\partial_{x_1}E_n^{(L-\hspace{-0.2mm}1)}\hspace{-0.2mm}(x\hspace{-0.1mm},\hspace{-0.2mm}x_{1:M}\hspace{-0.2mm},n_{1:M},\hspace{-0.2mm}N_{1:P},\hspace{-0.2mm}e)\hspace{-0.7mm}\Big)
\hspace{-0.8mm}\bigg]\hspace{-0.5mm}\raisebox{-1.0mm}{\Bigg|}\raisebox{0mm}{\hspace{-1.0mm}
\begin{minipage}[t]{57mm}
$\ss\hspace{-0.5mm} E_n^{(1)}=E_n^{(1)}(\hspace{-0.1mm}x'\hspace{-0.9mm},\hspace{0.3mm}x_{1:M}\hspace{-0.4mm},\hspace{0.2mm}n_{1:M}\hspace{-0.4mm},\hspace{0.1mm}e)$\\[-1.2mm]
$\ss\hspace*{-0.5mm} E_n^{(l)}\hspace{0.3mm}=E_n^{(l)}(\hspace{-0.1mm}x'\hspace{-0.9mm},\hspace{0.3mm}x_{1:M}\hspace{-0.4mm},\hspace{0.2mm}n_{1:M}\hspace{-0.4mm},\hspace{0.1mm}N_{1:P},\hspace{0.1mm}e),\,
l=2,\hspace{-0.7mm}...,\hspace{0.3mm}L-\hspace{-0.2mm}1$\\[-2mm]
$\ss x_1=\cdots=\hspace{0.3mm}x_M=\hspace{0.3mm}x$
\end{minipage}},
\end{eqnarray}
\end{linenomath}
\end{subequations}
whereas
\begin{linenomath}
\begin{eqnarray}
\label{eq:dEN}
& & \hspace{-12.0mm}
\Frac{\raisebox{-0.8mm}{$1$}}{\raisebox{1.0mm}{$n_1\hspace{-1.0mm}$}}
\hspace{-0.5mm}\left.\partial_{x_1^{\phantom{d}}}\hspace{-0.2mm}
E_N(x_{1:M},n_{1:M},\hspace{-0.2mm}N_{1:P},e)\right|_{x_1=\cdots=x_M=x}\hspace{-0.5mm}=
\Frac{\raisebox{-0.8mm}{$1$}}{\raisebox{1.0mm}{$n_1\hspace{-1.0mm}$}}
\bigg[
\partial_{x_1}e_N(x_1,\hspace{-0.3mm}E_n,e)\hspace{0.2mm}n_1\nonumber\\
& & \hspace{-6.0mm}+\,
\partial_{E_n^{(1)}}\hspace{0.0mm}e_N(x,\hspace{-0.3mm}E_n^{(1)},\hspace{-0.7mm}...,\hspace{-0.3mm}E_n^{(L)},e)\nonumber\\[-1.5mm]
& & \hspace{0.0mm}
\hspace{-0.5mm}\Big(\hspace{-0.5mm}
\partial_{x_1}E_n^{(1)}(x_1,\hspace{-0.5mm}[x,\hspace{-0.7mm}...,\hspace{-0.3mm}x],n_{1:M},\hspace{-0.3mm}e)\hspace{0.2mm}n_1
\hspace{-0.5mm}+\hspace{-0.5mm}\makebox[5.5mm][l]{$\sum\limits_{j=1}^M$}
\partial_{x_1}E_n^{(1)}(x\hspace{-0.1mm},\hspace{-0.2mm}x_{1:M}\hspace{-0.2mm},n_{1:M},\hspace{-0.2mm}e)\hspace{0.5mm}n_j\hspace{-0.5mm}\Big)\nonumber\\[-4mm]
& & \hspace{-5.0mm}
\vdots\nonumber\\[-1mm]
& & \hspace{-6.0mm}+\,
\partial_{E_n^{(L)}}e_N(x,\hspace{-0.3mm}E_n^{(1)},\hspace{-0.7mm}...,\hspace{-0.3mm}E_n^{(L)},e)
\hspace{-0.0mm}\Big(\hspace{-0.7mm}
\partial_{x_1}\hspace{-0.2mm}E_n^{(L)}\hspace{-0.1mm}(x_1\hspace{-0.2mm},\hspace{-0.7mm}[x,\hspace{-0.7mm}...,\hspace{-0.3mm}x],n_{1:M},\hspace{-0.2mm}N_{1:P},\hspace{-0.5mm}e)\hspace{0.2mm}n_1\nonumber\\
& & + \makebox[5.5mm][l]{$\sum\limits_{j=1}^M$}
\partial_{x_1}\hspace{-0.3mm}E_n^{(L)}\hspace{-0.1mm}(x\hspace{-0.1mm},\hspace{-0.2mm}x_{1:M}\hspace{-0.2mm},n_{1:M},\hspace{-0.2mm}N_{1:P},e)\hspace{0.2mm}n_j
\hspace{-0.6mm}\Big)
\hspace{-0.8mm}\bigg]\hspace{-0.5mm}\raisebox{-1.0mm}{\Bigg|}\raisebox{0mm}{\hspace{-1.0mm}
\begin{minipage}[t]{54mm}
$\ss\hspace{-0.5mm} E_n^{(1)}=E_n^{(1)}(\hspace{-0.1mm}x'\hspace{-0.9mm},\hspace{0.3mm}x_{1:M}\hspace{-0.4mm},\hspace{0.2mm}n_{1:M}\hspace{-0.4mm},\hspace{0.1mm}e)$\\[-1.2mm]
$\ss\hspace*{-0.5mm} E_n^{(l)}\hspace{0.3mm}=E_n^{(l)}(\hspace{-0.1mm}x'\hspace{-0.9mm},\hspace{0.3mm}x_{1:M}\hspace{-0.4mm},\hspace{0.2mm}n_{1:M}\hspace{-0.4mm},\hspace{0.1mm}N_{1:P},\hspace{0.1mm}e),\,
l=2,\hspace{-0.7mm}...,\hspace{0.3mm}L$\\[-2mm]
$\ss x_1=\cdots=\hspace{0.3mm}x_M=\hspace{0.3mm}x$
\end{minipage}}\nonumber\\
& & \hspace{-12.0mm}=
\bigg[
\partial_{x_1}e_N(x_1,\hspace{-0.3mm}E_n,e)\hspace{0.5mm}\nonumber\\
& & \hspace{-6.0mm}+\hspace{0.5mm}n\hspace{0.5mm}
\partial_{E_n^{(1)}}\hspace{0.0mm}e_N(x,\hspace{-0.3mm}E_n^{(1)},\hspace{-0.7mm}...,\hspace{-0.3mm}E_n^{(L)},e)\nonumber\\[0mm]
& & \hspace{0.0mm}
\hspace{-0.5mm}\Big(\hspace{-0.5mm}
\partial_{x_1}e_1(x_1,x,e)
\hspace{-0.5mm}+\hspace{-0.5mm}
\Frac{\raisebox{-0.8mm}{$1$}}{\raisebox{1.0mm}{$n_1\hspace{-1.0mm}$}}
\hspace{0.5mm}
\partial_{x_1}E_n^{(1)}(x\hspace{-0.1mm},\hspace{-0.2mm}x_{1:M}\hspace{-0.2mm},n_{1:M},\hspace{-0.2mm}e)
\hspace{-0.5mm}\Big)\nonumber\\[-4mm]
& & \hspace{-5.0mm}
\vdots\nonumber\\[-1mm]
& & \hspace{-6.0mm}+\hspace{0.5mm}n\hspace{0.5mm}
\partial_{E_n^{(L)}}\hspace{0.0mm}e_N(x,\hspace{-0.3mm}E_n^{(1)},\hspace{-0.7mm}...,\hspace{-0.3mm}E_n^{(L)},e)
\hspace{-0.5mm}\Big(\hspace{-0.5mm}
\partial_{x_1}e_L\hspace{-0.1mm}(x_1\hspace{-0.2mm},\hspace{-0.2mm}x,\hspace{-0.5mm}E_n^{(1)},\hspace{-0.7mm}...,\hspace{-0.5mm}E_n^{(L-\hspace{-0.2mm}1)},\hspace{-0.2mm}N_{1:P},e)
\hspace{0.0mm}\nonumber\\
& & \hspace{-0.0mm}+\hspace{0.5mm}
\Frac{\raisebox{-0.8mm}{$1$}}{\raisebox{1.0mm}{$n_1\hspace{-1.0mm}$}}
\hspace{0.5mm}
\partial_{x_1}E_n^{(L)}\hspace{-0.1mm}(x\hspace{-0.1mm},\hspace{-0.2mm}x_{1:M}\hspace{-0.2mm},n_{1:M},\hspace{-0.2mm}N_{1:P},\hspace{-0.2mm}e)\hspace{-0.7mm}\Big)
\hspace{-0.8mm}\bigg]\hspace{-0.5mm}\raisebox{-1.0mm}{\Bigg|}\raisebox{0mm}{\hspace{-1.0mm}
\begin{minipage}[t]{54mm}
$\ss\hspace{-0.5mm} E_n^{(1)}=E_n^{(1)}(\hspace{-0.1mm}x'\hspace{-0.9mm},\hspace{0.3mm}x_{1:M}\hspace{-0.4mm},\hspace{0.2mm}n_{1:M}\hspace{-0.4mm},\hspace{0.1mm}e)$\\[-1.2mm]
$\ss\hspace*{-0.5mm} E_n^{(l)}\hspace{0.3mm}=E_n^{(l)}(\hspace{-0.1mm}x'\hspace{-0.9mm},\hspace{0.3mm}x_{1:M}\hspace{-0.4mm},\hspace{0.2mm}n_{1:M}\hspace{-0.4mm},\hspace{0.1mm}N_{1:P},\hspace{0.1mm}e),\,
l=2,\hspace{-0.7mm}...,\hspace{0.3mm}L$\\[-2mm]
$\ss x_1=\cdots=\hspace{0.3mm}x_M=\hspace{0.3mm}x$
\end{minipage}}\hspace{-0.5mm}.
\end{eqnarray}
\end{linenomath}

Note that eq.~\eqref{eq:dEq} expresses the $E_n^{(L)}$-derivative w.r.t.\hspace{-0.5mm} $x_1$ in terms of $E_n^{(l)}$-ones with $l\hspace{-0.3mm}<L$ (including the $1/n_1$ factor in front), and should therefore be used recursively (for decreasing $L$) up to $L=1$, where \eqref{eq:dE1} then applies.
Once all recursions are taken and the resulting expressions substituted into \eqref{eq:dEN} and then into \eqref{eq:phi11_1}, the right hand side indeed becomes a function of $(n,\hspace{-0.3mm}N_{1:P},\hspace{-0.3mm}x,\hspace{-0.3mm}x')$, that is function $\phi_{1,1}$ in \eqref{eq:phi11}.
In fact, the densities $n_{1:M}$ appear in the right hand side of \eqref{eq:phi11_1} only through the environmental quantities in $E_n$ and $E_N$, and only the total density $n$ matters when the final evaluation at $x_1\hspace{-0.5mm}=\!\cdot\!\cdot\!\cdot\!=\hspace{-0.5mm}x_M\hspace{-0.5mm}=\hspace{-0.5mm}x$ is taken in eqs.~(\ref{eq:dEq},\hspace{0.2mm}\ref{eq:dEN}). 
Further note that the obtained $\phi_{1,1}$ is indeed independent on the number $M$ of considered morphs.

\section{Discussion and future directions}
\label{sec:dis}
In this paper I have developed a general procedure to formulate asexual (unstructured, deterministic) population models resulting from individual pairwise interactions and possibly characterized by nonlinear density dependencies in the per-capita growth rates (nonlinear fitnesses).

The models are underpinned with individual strategies that represent behavioral, morphological, and functional traits.
Individuals are assumed to interact in pairs and the effects of an interaction on the growth rates of the corresponding populations is determined by the strategies of the interacting individuals, other than by the current state of the biotic and abiotic environments---the first described by the densities of all populations in the community, the second defined by possibly fluctuating exogenous variables.
Intra- as well as inter-specific interactions are considered.

The procedure exploits the notions of fitness generating function \citep{Vincent05} and of environmental feedback \citep{Mylius_and_Diekmann_95_OI,Metz_and_Gyllenberg_01_PRSB}, namely the practice of writing the per-capita growth rate of a virtual population with strategy $x'$ (and zero density) in two steps.
First, a set of environmental quantities, possibly functions of $x'$, is defined based on the biotic and abiotic components of the environment.
Then, the growth rate is written as a function of the strategy $x'$ and of the environmental quantities evaluated at $x'$, which therefore act as a feedback from the current state of the environment to the rule---the growth rate---determining the (biotic) environment in the near future.
When the environmental quantities are all independent of $x'$, i.e., they are only affected by the resident strategies present in the community, the environment is said to be finite dimensional, with dimension given by the number of necessary quantities.
The environment is otherwise infinite dimensional and note that this is the situation whenever the strategies of two conspecifics jointly determine the result of their interaction (see e.g. prey competition in Box~3).

The environment is defined w.r.t.\hspace{-0.5mm} a focus species present in the community in $M\hspace{-0.2mm}\ge 1$ resident morphs, described by strategies $x_1,x_2,\hspace{-0.6mm}...,x_M$ and densities $n_1(t),n_2(t),\hspace{-0.6mm}...,\hspace{-0.2mm}n_M(t)$ at time $t$ (e.g. the prey species in Box~3).
The environment is composed of four sets of quantities:
$E_n^{(l)}$, $l=1,\hspace{-0.6mm}...,\hspace{-0.2mm}L$, typically functions of $x'$ and describing the intra-specific interactions of the $x'$-strategist;
$E_N$, $x'$-independent and describing the inter-specific interactions of the resident strategies $x_1,x_2,\hspace{-0.6mm}...,x_M$;
the densities $N_1(t),N_2(t),\hspace{-0.6mm}...,\hspace{-0.3mm}N_P(t)$ of the $P\ge 0$ populations of different species;
and the (possibly fluctuating) abiotic environment $e(t)$.

The environmental quantities in $E_n^{(l)}$ are defined by an iterative procedure (Sect.~\ref{ssec:proc}) that integrates a suitable interacting kernel $e_n^{(l)}$ over the impulsive density distribution $\nu$ (defined in eq.~\eqref{eq:nu}).
The interacting kernel $e_n^{(1)}$ is simply a functions of the interacting strategies $(x'\hspace{-0.5mm},x_j)$, $j=1,\hspace{-0.6mm}...,\hspace{-0.2mm}M$, and these are the ``first-level'' interactions (e.g. prey competition and predation in Box~3).
But more in general, the interacting kernel $e_n^{(L)}$ of level $L>1$ also depends on the environmental quantities $E_n^{(l)}$ of level $l<L$, characterizing the (intra-specific) activities of $x_j$-strategists.
In this way, ``second-level'' interactions ($L=2$) take into account that the encountered $x_j$-strategist is involved, at the same time, in other (intra-specific) first-level activities
(see the $g$-function in Box~4, second term, where the $x'$-strategist is cannibalized by $x_j$-individuals who at the same time harvest the common resource $n_0$ and cannibalize $x_k$-strategists).
Similarly, third- and higher-level ($L\hspace{-0.4mm}>\hspace{-0.4mm}2$) interacting kernels describe interactions that are affected by the concurrent lower-level activities of the involved resident individual
(though, to my knowledge, no model in the literature requires more than two levels).
Thus, other than the dimensionality of the environment---that is interesting because it sets, when finite, a limit on the number of stationarily coexisting morphs of the focus species \citep{Diekmann_et_al_03_TPB,Meszena_et_al_06_TPB}---the proposed procedure highlights another characterizing feature of the environmental feedback, i.e., the number $L$ of ``interacting levels.''
In other words, the $L\hspace{-0.2mm}-\hspace{-0.4mm}1$ is the the number of nested intra-specific interacting dependencies.

Note that \citet{Geritz05a} distinguishes between linear and nonlinear environments, which basically corresponds to the distinction $L=1$ and $L>1$, respectively.
In fact, while the environmental quantities in $E_n^{(1)}$ are necessarily linear in the densities of the resident strategies, nonlinearities may appear only through the dependency of an interacting kernel $e_n^{(L)}$ of level $L>1$ on a lower-level quantity $E_n^{(l)}$, $l<L$.
Thus, not all nonlinearities are allowed, only those that can be obtained through the proposed procedure.

The procedure also highlights the structural properties fulfilled by the population per-capita growth rates (Sect.~\ref{ssec:prop}).
While properties P1--3 are rather obvious, property P4 is more involved.
It basically generalizes the law of mass-action that, in the case of unstructured populations, only yields per-capita growth rates with linear density-dependence and coefficients that are functions of the strategies of the two interacting individuals (see Box~1).
Thus, property P4 under strict mass-action simply says that, when all strategies are identical, pure fitness derivatives w.r.t.\hspace{-0.5mm} a resident strategy are proportional to the corresponding density (through a density-independent proportionality coefficient) and all mixed derivatives are null.
But when fitnesses are nonlinear, terms proportional to powers of the densities of the perturbed strategies appear, in accordance to the scheme described in Sect.~\ref{ssec:prop}.
Note that strict mass-action implies a linear environment ($L=1$), though a nonlinear fitness might arise in one- as well as multi-level environments (see Box~3 and~4, respectively).

Note also that property P4 can be further generalized.
In fact, if only a subset  $x_1,x_2,\hspace{-0.6mm}...,x_m$, $m\hspace{-0.8mm}<\hspace{-0.5mm}M$, of the resident strategies are identical to $x$, then, when perturbing $k$ of those strategies, the functions $\phi$'s and $\psi$'s take the sum \mbox{$n_1\!+\!\cdot\!\cdot\!\cdot\!+\!n_m$} and the other densities $n_{m+1},\hspace{-0.6mm}...,n_M$, as well as the strategy $x$ and the other strategies $x_{m+1},\hspace{-0.6mm}...,x_M$, as independent arguments.
Moreover, properties P1--4 are not necessarily due to the pairwise nature of the allowed interactions.
In a modeling framework where $k$-wise group interactions, $k\hspace{-0.5mm}\ge\hspace{-0.5mm}3$, are relevant, they could be accounted for by an iterative procedure (just like that in Sect.~\ref{ssec:proc}) with $(k\hspace{-0.5mm}-\hspace{-0.5mm}1)$-dimensional integrals, and properties P1--4 would still follow, though their derivation from the procedure would be more involved.

The structure of the fitness derivatives w.r.t.\hspace{-0.5mm} the resident strategies have already been considered by \citet{Meszena05a}, who define the per-capita growth rate of a virtual strategy $x'$ in a single-species polymorphic community as an $x'$-dependent functional $r$ over the distribution $\nu$.
Taking the derivative of $r(x'\hspace{-0.5mm},\nu)$ w.r.t.\hspace{-0.5mm} a resident strategy $x_i$ then requires a notion for the functional derivative of $r$ w.r.t.\hspace{-0.5mm} the distribution $\nu$ that preserves the chain rule---a concept defined by \citet{Meszena05a} in Appendix.
Exploiting this concept, they were able to show that the fitness derivative w.r.t.\hspace{-0.5mm} a resident strategy is proportional to the density of the perturbed strategy with a proportionality coefficient that is density-dependent only through the total density of the species, i.e., property P4 for $k=1$ and $d_1=1$ (that basically follows from strict mass-action).
My approach therefore extends the one in \citet{Meszena05a} and avoids the complicacies of functional analysis.
I assume that the functional $r$ is obtained through the iterative procedure described in Sect.~\ref{ssec:proc}, which is ultimately the only way to describe pairwise interactions by smoothly operating on the distribution $\nu$.

The presented approach (as well as the one in \citealp{Meszena05a}) is restricted to the case of unstructured populations, i.e., to cases in which the population abundances can be considered as complete descriptions of the populations' state
\citep[this is often justified by the time-scale separation argument used in Box~2, as the dynamics of the population structure can be considered at equilibrium on the slower time-scale of ecology, see e.g.][]{Greiner_et_al_94_CAMQ}.
However, it is reasonable to expect the approach to be extendable, with the proper mathematical expertise, to the generality of structured populations (at least to the case of a finite number of classes).


I see property P4 as the main contribution of this work, rather than the iterative procedure of Sect.~\ref{ssec:proc}.
The procedure is important because it implies property P4, thus justifying its assumption when considering a generic $g$-function, instead of a specific one derived through the procedure.
The procedure also implies properties P1--3, though these are more obvious to be assumed.
From a formal point of view, it would be interesting to prove that properties P1--4 all together imply that the $g$-function can be obtained through the procedure.
However, in practice, it is enough to know that properties P1--4 hold by construction of the model.
Indeed, besides the theoretical interest, property P4 has relevant implications when studying the competition between similar strategies.

For example, by setting $x_i:=x+\eps\cos\th_i$, $\sum_{i=1}^M\cos^2\th_i=1$, close to a reference strategy $x$, it is possible to express the $g$-derivatives w.r.t.\hspace{-0.5mm} $\eps$ at $\eps=0$ in terms of the derivatives of the monomorphic $g$-function, i.e.,
\begin{linenomath}
\begin{eqnarray*}
 & & \hspace{-2mm}\partial_{\eps}\hspace{0.2mm}g_M(n_{1:M},\hspace{-0.3mm}N_{1:P},x_{1:M},x'\hspace{-0.5mm},t)\Big|_{\eps=0}\\[-2mm]
 & & \hspace{6mm}=\hspace{-0.5mm}
\sum_{i=1}^{M}\hspace{-0.3mm}\phi_{1,1}(n,\hspace{-0.3mm}N_{1:P},x,x'\hspace{-0.5mm},t)\hspace{0.2mm} r_i\,n\cos\th_i =
\partial_x\hspace{0.5mm}g_1(n,\hspace{-0.3mm}N_{1:P},x,x'\hspace{-0.5mm},t)\hspace{-0.5mm}\sum_{i=1}^{M}\hspace{-0.3mm}r_i\cos\th_i,
\end{eqnarray*}
\end{linenomath}
where $r_i:=n_i/n$ is the relative densities of the resident strategy $x_i$.
Thus, the radial expansion (w.r.t.\hspace{-0.5mm} $\eps$ along the ray defined by the angles $\th_i$, $i=1,\hspace{-0.6mm}...,\hspace{-0.2mm}M$) of a generic $M$-morphic $g$-function can be characterized in terms of the monomorphic $g$-function at the reference strategy $x$.
This allows to rewrite the ecological model \eqref{eq:eco1} using the total density $n$ and the relative densities $r_i$ as new variables.
The total density $n$ (as well as the densities $N_{1:P}$ of the other species) converges to the monomorphic ecological equilibrium $\bar{n}(x)$ ($\bar{N}_{1:P}(x)$) much faster than the relative dynamics between the various morphs, which can then be studied in isolation, thanks to the time-scale separation, and characterized in terms of the properties of the monomorphic fitness.
This analysis has been carried out by \citet{Meszena05a} up to first-order in $\eps$ to characterize the competition between similar strategies close to the singular strategies of adaptive dynamics \citep{Geritz97,Geritz98,Dercole_and_Rinaldi_08_PUP}.
Particularly interesting is the extension to the higher-orders for $M=2$, to characterize the resident-invader dynamics close to degenerate singular strategies. It is however out of the scope of this paper and will be presented elsewhere.

Switching to evolutionary considerations, after the coexistence of a resident and an invader morphs, say with strategies $x_1$ and $x_2$ close to a singular strategy $x$, the subsequent evolutionary dynamics are ruled by the geometry of the dimorphic invasion fitness
\begin{linenomath}
$$
\lambda_2(x_1,x_2,x')=g_2(\bar{n}_1(x_1,x_2),\bar{n}_2(x_1,x_2),\bar{N}_{1:P}(x_1,x_2),x_1,x_2,x')
$$
\end{linenomath}
\citep{Metz92}, where $\bar{n}_1(x_1,x_2)$, $\bar{n}_2(x_1,x_2)$, and $\bar{N}_{1:P}(x_1,x_2)$ are the equilibrium densities of ecological coexistence and a time-invariant abiotic environment has been assumed.
As it will be discussed in another contribution, the dimorphic invasion fitness is nonsmooth and cannot be expanded w.r.t.\hspace{-0.5mm} $(x_1,x_2)$ close to $(x,x)$.
The expansion must be taken along radial directions in the strategy space $(x_1,x_2)$, i.e., w.r.t.\hspace{-0.5mm} $\eps$ with $x_1=x+\eps\cos\th$ and $x_2=x+\eps\sin\th$.
Then, by property P4, the $\eps$-expansion of the dimorphic fitness can be written in terms of the derivatives of the monomorphic one
\begin{linenomath}
$$
\lambda_1(x,x')=g_1(\bar{n}(x),\bar{N}_{1:P}(x),x,x').
$$
\end{linenomath}
Specifically, a sort of normal form approach, borrowed from bifurcation theory, should become possible to fully characterize the adaptive dynamics after evolutionary branching in terms of invasion criteria to be evaluated before branching.

In conclusion, the proposed procedure for building asexual (unstructured, deterministic) ecological models, and property P4 in particular, open up new directions along which the theoretical aspects of ecology and evolutionary biology can be investigated.

\section*{Acknowledgements}
I am indebted with two young collaborators of mine, Fabio Della Rossa and Pietro Landi, and with Stefan Geritz and Hans Metz, from Helsinki and Leiden Universities, for many fruitful discussions.
This work was supported mainly by the Italian Ministry for University and Research (under contract FIRB RBFR08TIA4).

\newpage
{\small
\section*{Box 1}
\renewcommand{\theequation}{B1.\arabic{equation}}
\setcounter{equation}{0}  
The equivalent chemical reactions (left) and the ODEs (right) of the Lotka-Volterra $M$-prey-one-predator model ($i,j=1,\hspace{-0.6mm}...,\hspace{-0.3mm}M$):\\
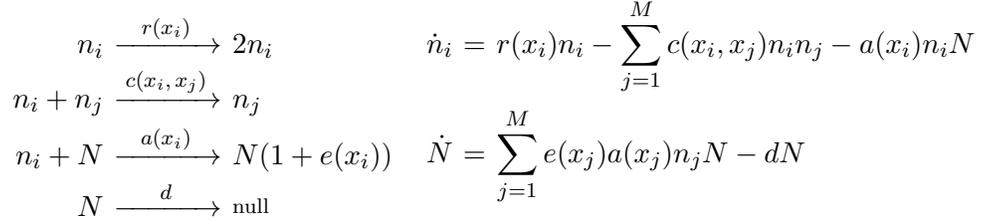
\begin{minipage}[t]{55mm}
\vspace*{-4mm}
\begin{linenomath}
\begin{eqnarray*}
\makebox[8mm][r]{$n_i$} & \xrightarrow{\makebox[12mm][c]{\ssz $r(x_i)$}} & 2n_i\\
\makebox[8mm][r]{$n_i + n_j$} & \xrightarrow{\makebox[12mm][c]{\ssz $c(x_i,x_j)$}} & n_j\\
\makebox[8mm][r]{$n_i + N$} & \xrightarrow{\makebox[12mm][c]{\ssz $a(x_i)$}} & \makebox[12mm][l]{$N(1+e(x_i))$}\\
\makebox[8mm][r]{$N$} & \xrightarrow{\makebox[12mm][c]{\ssz $d$}} & \raisebox{0.4mm}{\ssz null}
\end{eqnarray*}
\end{linenomath}
\end{minipage}
\begin{minipage}[t]{60mm}
\vspace*{-6.2mm}
\begin{linenomath}
\begin{eqnarray*}
\makebox[5mm][r]{$\dot{n}_i$} & = & \makebox[45mm][l]{$\ds r(x_i)n_i -\sum_{j=1}^M c(x_i,x_j)n_in_j-a(x_i)n_iN$}\\[0.5mm]
\makebox[5mm][r]{$\dot{N}$} & = & \sum_{j=1}^M e(x_j)a(x_j)n_jN -dN
\end{eqnarray*}
\end{linenomath}
\end{minipage}\\[3.5mm]
The translation of the chemical reactions into the ODEs for the population densities follows the rule\\[-2mm]
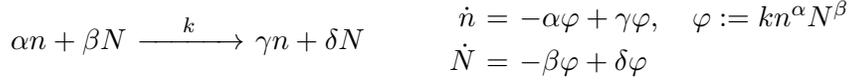
\begin{minipage}{55mm}
\begin{linenomath}
\begin{eqnarray*}
\makebox[8mm][r]{$\alpha n + \beta N$} & \xrightarrow{\makebox[12mm][c]{\ssz $k$}} & \makebox[12mm][l]{$\gamma n + \delta N$}
\end{eqnarray*}
\end{linenomath}
\end{minipage}
\begin{minipage}{60mm}
\vspace*{-0.3mm}
\begin{linenomath}
\begin{eqnarray*}
\makebox[5mm][r]{$\dot{n}$} & = & \makebox[45mm][l]{$-\alpha\varphi + \gamma\varphi,\quad \varphi:=kn^\alpha N^\beta$}\\
\makebox[5mm][r]{$\dot{N}$} & = & -\beta\varphi + \delta\varphi
\end{eqnarray*}
\end{linenomath}
\end{minipage}\\[2mm]
where:
\begin{itemize}
\vspace*{-2mm}\item[--]
$n$ and $N$ are the densities of the reacting populations;
\vspace*{-2mm}\item[--]
$\alpha$ and $\beta$ are nonnegative integers such that $1\le \alpha+\beta\le 2$, i.e., one or two molecules are reacting;
\vspace*{-2mm}\item[--]
$\gamma$ and $\delta$ are nonnegative coefficients defining the products of the reaction;
\vspace*{-2mm}\item[--]
$k$ is the kinetic rate of the reaction;
\vspace*{-2mm}\item[--]
$\varphi$ is the encounter rate, i.e., the number of encounters of the type $\alpha n + \beta N$ per unit of time;
\vspace*{-2mm}\item[--]
$\alpha\varphi$ (resp. $\beta\varphi$) molecules of $n$ (resp. $N$) are consumed per unit of time (and of space), whereas $\gamma\varphi$ (resp. $\delta\varphi$) are produced, the balance giving the dynamics of $n$ (resp. $N$).
\end{itemize}
The chemical reactions, from top to bottom, describe:
\begin{itemize}
\vspace*{-2mm}\item[--]
prey reproduction, one individual producing a second one at rate $r(x_i)=\text{birth rate}-\text{death rate}$;
\vspace*{-2mm}\item[--]
prey mortality due to intra-specific competition, individuals $i$ and $j$ (resp. of populations $n_i$ and $n_j$) are involved in a contest, $j$ survives;
\vspace*{-2mm}\item[--]
predation of a prey individual of population $n_i$ and predator reproduction with prey-to-predator conversion factor $e(x_i)$;
\vspace*{-2mm}\item[--]
predator mortality at rate $d$.
\end{itemize}}

\newpage
{\small
\section*{Box 2}
\renewcommand{\theequation}{B2.\arabic{equation}}
\setcounter{equation}{0}  
The equivalent chemical reactions (left) and the ODEs (right) of the behaviorally structured $M$-prey-one-predator model ($i,j=1,\hspace{-0.6mm}...,\hspace{-0.3mm}M$):\\\begin{minipage}[t]{55mm}
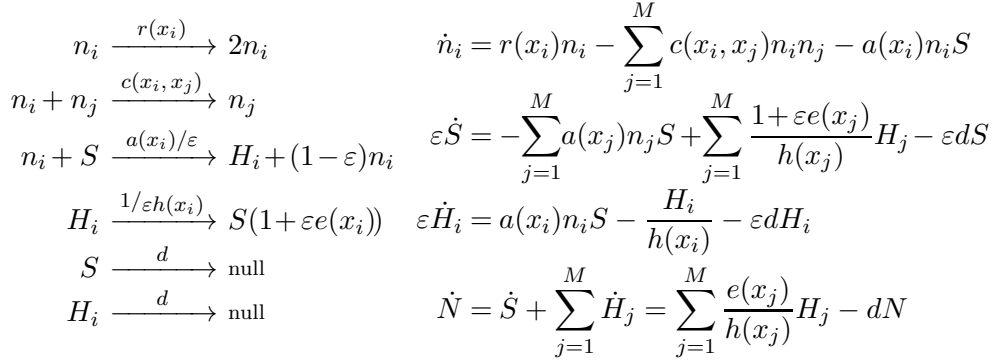
\vspace*{-4mm}
\begin{linenomath}
\begin{eqnarray*}
\makebox[0mm][r]{$n_i$} & \xrightarrow{\makebox[12mm][c]{\ssz $r(x_i)$}} & 2n_i\\
\makebox[0mm][r]{$n_i\hspace{-0.3mm} + n_j$} & \xrightarrow{\makebox[12mm][c]{\ssz $c(x_i,x_j)$}} & n_j\\
\makebox[0mm][r]{$n_i\hspace{-0.3mm} + S$} & \xrightarrow{\makebox[12mm][c]{\ssz $a(x_i)/\eps$}} &
\makebox[12mm][l]{$H_i\hspace{-0.5mm}+\hspace{-0.5mm}(1\hspace{-0.5mm}-\hspace{-0.3mm}\eps)n_i$}\\[0.7mm]
\makebox[0mm][r]{$H_i$} & \xrightarrow{\makebox[12mm][c]{\ssz $1/\raisebox{-0.5mm}{$\eps h(x_i)$}$}} & \makebox[12mm][l]{$S(1\hspace{-0.7mm}+\hspace{-0.3mm}\eps e(x_i)\hspace{-0.5mm})$}\\[-0.5mm]
\makebox[0mm][r]{$S$} & \xrightarrow{\makebox[12mm][c]{\ssz $d$}} & \raisebox{0.4mm}{\ssz null}\\
\makebox[0mm][r]{$H_i$} & \xrightarrow{\makebox[12mm][c]{\ssz $d$}} & \raisebox{0.4mm}{\ssz null}
\end{eqnarray*}
\end{linenomath}
\end{minipage}
\begin{minipage}[t]{60mm}
\vspace*{-6.2mm}
\begin{linenomath}
\begin{eqnarray*}
\makebox[5mm][r]{$\dot{n}_i$} & \hspace{-0.5mm}=\hspace{-0.5mm} &
r(x_i)n_i - \sum_{j=1}^M c(x_i,x_j)n_in_j-a(x_i)n_iS\\[-2mm]
\makebox[5mm][r]{$\eps\dot{S}$} & \hspace{-0.5mm}=\hspace{-0.5mm} & \makebox[45mm][l]{$\ds
-\hspace{-0.8mm}\sum_{j=1}^M\hspace{-0.8mm}a(x_j)n_jS\hspace{-0.3mm}+\hspace{-1.0mm}\sum_{j=1}^M \Frac{1\hspace{-0.7mm}+\hspace{-0.3mm}\eps e(x_j)}{h(x_j)}H_j\hspace{-0.5mm}-\eps dS$}\\[-1.0mm]
\makebox[5mm][r]{$\eps\dot{H}_i$} & \hspace{-0.5mm}=\hspace{-0.5mm} &
a(x_i)n_iS-\Frac{H_i}{h(x_i)}-\eps dH_i\\[-0.2mm]
\makebox[5mm][r]{$\dot{N}$} & \hspace{-0.5mm}=\hspace{-0.5mm} &
\dot{S}+\sum_{j=1}^M\dot{H}_j = \sum_{j=1}^M\Frac{e(x_j)}{h(x_j)}H_j-dN
\end{eqnarray*}
\end{linenomath}
\end{minipage}\\[2.0mm]
The predator population is partitioned into the searching and handling subpopulations $N=S+\sum_{j=1}^M H_j$.
Prey capturing and handling are assumed to be fast processes ($\eps$ small) compared with prey reproduction, competition, and predator mortality.
At each prey-predator encounter, a fraction $\eps$ of prey is consumed and the corresponding amount of predator newborn is $\eps e(x_i)$.
The Holling-type-II functional response is obtained in the limit $\eps\to 0$ of the time-scale separation, yielding the fast equilibrium
\begin{linenomath}
$$
\bar{S}=N\bigg(\hspace{-0.5mm}1\hspace{-0.5mm}+\hspace{-0.5mm}\sum_{j=1}^M a(x_j)h(x_j)n_j\hspace{-0.5mm}\bigg)^{\hspace{-1.0mm}-1}\quad
\bar{H}_i=a(x_i)h(x_i)n_iN\bigg(\hspace{-0.5mm}1\hspace{-0.5mm}+\hspace{-0.5mm}\sum_{j=1}^M a(x_j)h(x_j)n_j\hspace{-0.5mm}\bigg)^{\hspace{-1.0mm}-1}
$$
\end{linenomath}
and the following slow unstructured equations (after substituting $S$ and $H_i$ with $\bar{S}$ and $\bar{H}_i$):
\begin{linenomath}
\begin{subequations}
\label{eq:HII}
\begin{eqnarray}
\label{eq:HIIa}
\hspace{-10.0mm} \dot{n}_i & \hspace{-0.5mm}=\hspace{-0.5mm} & r(x_i)n_i -\sum_{j=1}^M c(x_i,x_j)n_in_j-a(x_i)n_iN
\bigg(\hspace{-0.5mm}1\hspace{-0.5mm}+\hspace{-0.5mm}\sum_{j=1}^M a(x_j)h(x_j)n_j\hspace{-0.5mm}\bigg)^{\hspace{-1.0mm}-1}\\
\label{eq:HIIb}
\hspace{-10.0mm} \dot{N} & \hspace{-0.5mm}=\hspace{-0.5mm} &
\sum_{j=1}^M e(x_j)a(x_i)n_jN
\bigg(\hspace{-0.5mm}1\hspace{-0.5mm}+\hspace{-0.5mm}\sum_{j=1}^M a(x_j)h(x_j)n_j\hspace{-0.5mm}\bigg)^{\hspace{-1.0mm}-1}
\hspace{-1.0mm}-\hspace{1.0mm}dN
\end{eqnarray}
\end{subequations}
\end{linenomath}}

\newpage
{\small
\section*{Box 3}
\renewcommand{\theequation}{B3.\arabic{equation}}
\setcounter{equation}{0}  
Direct formulation of the nonlinear fitness for the Holling-type-II $M$-prey-one-predator model.
The prey $g$-function for model \eqref{eq:HII} is time-invariant and reads:
\begin{linenomath}
\begin{eqnarray*}
g_M(n_{1:M},N,x_{1:M},x') & \hspace{-0.5mm}:=\hspace{-0.2mm} &
r(x') \hspace{-0.2mm}-\hspace{-0.8mm}\sum_{j=1}^M\hspace{-0.5mm} c(x',x_j)n_j \hspace{-0.4mm}-\hspace{-0.3mm} a(x')\hspace{0.1mm}N\hspace{-0.2mm}
\bigg(\hspace{-0.5mm}1\hspace{-0.5mm}+\hspace{-0.5mm}\sum_{j=1}^M a(x_j)h(x_j)n_j\hspace{-0.5mm}\bigg)^{\hspace{-1.0mm}-1}\hspace{-0.7mm}.
\end{eqnarray*}
\end{linenomath}
It can be constructed with the procedure of Sect.~\ref{ssec:proc} in one step ($L=1$) with the following time-invariant interacting kernels:
\begin{linenomath}
$$
e_n^{(1)}(x'\hspace{-0.5mm},x):=c(x'\hspace{-0.5mm},x) \quad\text{and}\quad
e_{N,1}(x):=a(x)h(x)
$$
\end{linenomath}
(the subscript $1$ indicates the first component of $e_N$, a second is needed later) yielding
\begin{linenomath}
$$
E_n^{(1)}(x'\hspace{-0.5mm},x_{1:M},n_{1:M})=\sum_{j=1}^M\hspace{-0.5mm} c(x',x_j)n_j\quad\text{and}\quad
E_{N,1}(x_{1:M},n_{1:M})=\sum_{j=1}^M a(x_j)h(x_j)n_j,
$$
\end{linenomath}
and with the following $g$-function:
\begin{linenomath}
\begin{multline*}
g(x'\hspace{-0.5mm},E_n(x'\hspace{-0.5mm},x_{1:M},n_{1:M}),E_N(x_{1:M},n_{1:M}),N):=\\
r(x')-\hspace{-0.2mm}E_n^{(1)}(x'\hspace{-0.5mm},x_{1:M},n_{1:M})-
\Frac{a(x')N}{1\hspace{-0.5mm}+\hspace{-0.5mm}E_{N,1}(x_{1:M},n_{1:M})}
\end{multline*}
\end{linenomath}
(the abiotic environmental factors are constant parameters that shape the demographic functions $r$, $c$, $a$, and $h$, so the argument $e$ is omitted in the formulation).
One step is sufficient because intra-specific interactions only occur through competition and the competition coefficient $c(x',x_j)$ does not depend on the concomitant activities of the involved $x_j$-strategist.
Note that $g$ does not depend directly on the resident prey densities $n_1,\hspace{-0.6mm}...,n_M$, but the density dependence indirectly occurs through the environmental quantities in $E_n^{(1)}$ and $E_{N,1}$.

Taking the first and second derivatives of $g_2$ w.r.t.\hspace{-0.5mm} $x_1$, replacing whenever possible $n_1\hspace{-0.6mm}+\hspace{-0.3mm}n_2$ with $n$ (i.e., replacing $n_2$ with $n\hspace{-0.3mm}-\hspace{-0.3mm}n_1$ and simplifying the resulting expression), and collecting the remaining powers of $n_1$, one gets (see Supplementary Material):
\begin{linenomath}
\begin{eqnarray*}
\left.\partial_{x_1^{\phantom{2}}}\hspace{0.2mm}
g_2(n_1,n_2,N,x_1,x_2,x')\right|_{x_1=x_2=x} & \hspace{-0.5mm}=\hspace{-0.2mm} &
\phi_{1,1}(n,N,x,x')\hspace{0.2mm}n_1,\\
\left.\partial_{x_1^2}\hspace{0.2mm}
g_2(n_1,n_2,N,x_1,x_2,x')\right|_{x_1=x_2=x} & \hspace{-0.5mm}=\hspace{-0.2mm} &
\phi_{2,1}(n,N,x,x')\hspace{0.2mm}n_1 + \phi_{2,2}(n,N,x,x')\hspace{0.2mm}n_1^2,
\end{eqnarray*}
\end{linenomath}
with
\begin{linenomath}
\begin{eqnarray*}
\phi_{1,1}(n,\hspace{-0.7mm}N\hspace{-0.7mm},\hspace{-0.3mm}x,\hspace{-0.5mm}x')
 & \hspace{-0.7mm}=\hspace{-0.5mm} &
\frac{a(x')\hspace{-0.8mm}\left(\partial_x a(x)h(x)+a(x)\partial_x h(x)\right)}{(1+a(x)h(x)n)^2}N
\hspace{-0.5mm}-\partial_x c(x'\hspace{-0.5mm},x),\\[1.0mm]
\phi_{2,1}(n,\hspace{-0.7mm}N\hspace{-0.7mm},\hspace{-0.3mm}x,\hspace{-0.5mm}x')
 & \hspace{-0.7mm}=\hspace{-0.5mm} &
\frac{a(x')\hspace{-0.8mm}\left(\hspace{-0.3mm}\partial_{x^2} a(x)h(x)
\hspace{-0.8mm}+\hspace{-0.7mm} 2\partial_x a(x) \partial_x h(x)
\hspace{-0.8mm}+\hspace{-0.7mm} a(x)\partial_{x^2} h(x)\hspace{-0.4mm}\right)\hspace{-0.5mm}}{(1+a(x)h(x)n)^2}N
\hspace{-1.0mm}-\hspace{-0.3mm}\partial_{x^2}c(x'\hspace{-0.5mm},x),\\
\phi_{2,2}(n,\hspace{-0.7mm}N\hspace{-0.7mm},\hspace{-0.3mm}x,\hspace{-0.5mm}x')
 & \hspace{-0.7mm}=\hspace{-0.5mm} &
-\frac{2a(x')\hspace{-0.8mm}\left(\partial_x a(x)h(x)+a(x)\partial_x h(x)\right)^2}{(1+a(x)h(x)n)^3}N.
\end{eqnarray*}
\end{linenomath}
Taking the second mixed derivative of $g_3$ w.r.t.\hspace{-0.5mm} $x_{1,2}$, replacing replacing $n_3$ with $n\hspace{-0.3mm}-\hspace{-0.3mm}n_1\hspace{-0.3mm}-\hspace{-0.3mm}n_2$, simplifying, and collecting the monomial $n_1n_2$, one gets:
\begin{linenomath}
$$
\left.\partial_{x_1,x_2}\hspace{0.2mm}
g_3(n_1,n_2,n_3,N,x_1,x_2,x_3,x')\right|_{x_1=x_2=x_3=x} =
\phi_{1,1,1,1}(n,N,x,x')\hspace{0.2mm}n_1n_2,
$$
\end{linenomath}
with $\phi_{1,1,1,1}=\phi_{2,2}$ as expected from \eqref{eq:phi1111}.

Similarly, from \eqref{eq:HIIb}, the predator population growth rate reads:
\begin{linenomath}
$$
F_M(n_{1:M},N,x_{1:M}):=\sum_{j=1}^M e(x_j)a(x_j)n_jN
\bigg(\hspace{-0.5mm}1\hspace{-0.5mm}+\hspace{-0.5mm}\sum_{j=1}^M a(x_j)h(x_j)n_j\hspace{-0.5mm}\bigg)^{\hspace{-1.0mm}-1}
\hspace{-1.5mm}-\hspace{1.0mm}dN.
$$
\end{linenomath}
It can be constructed with the procedure of Sect.~\ref{ssec:proc} with the further interacting kernel
\begin{linenomath}
$$
e_{N,2}(x):=e(x)a(x),\quad\text{yielding}\quad E_{N,2}(x_{1:M},n_{1:M})=\sum_{j=1}^M e(x_j)a(x_j)n_j,
$$
\end{linenomath}
and with the following $F$-function:
\begin{linenomath}
$$
F(E_N(x_{1:M},n_{1:M}),N)=
\Frac{E_{N,2}(x_{1:M},n_{1:M})}
{1\hspace{-0.5mm}+\hspace{-0.5mm} E_{N,1}(x_{1:M},n_{1:M})}N - dN.
$$
\end{linenomath}

Taking the first and second derivatives of $F_2$ w.r.t.\hspace{-0.5mm} $x_1$, replacing $n_2$ with $n\hspace{-0.3mm}-\hspace{-0.3mm}n_1$, simplifying, and collecting $n_1$-powers, one gets (see Supplementary Material):
\begin{linenomath}
\begin{eqnarray*}
\left.\partial_{x_1^{\phantom{2}}}\hspace{0.2mm}
F_2(n_1,n_2,N,x_1,x_2)\right|_{x_1=x_2=x} & \hspace{-0.5mm}=\hspace{-0.2mm} &
\psi_{1,1}(n,N,x)\hspace{0.2mm}n_1,\\
\left.\partial_{x_1^2}\hspace{0.2mm}
F_2(n_1,n_2,N,x_1,x_2)\right|_{x_1=x_2=x} & \hspace{-0.5mm}=\hspace{-0.2mm} &
\psi_{2,1}(n,N,x)\hspace{0.2mm}n_1 + \psi_{2,2}(n,N,x)\hspace{0.2mm}n_1^2,
\end{eqnarray*}
\end{linenomath}
with
\begin{linenomath}
\begin{eqnarray*}
\psi_{1,1}(n,\hspace{-0.7mm}N\hspace{-0.7mm},\hspace{-0.3mm}x)
 & \hspace{-0.7mm}=\hspace{-0.5mm} &
\frac{\partial_x e(x)a(x)}{1+a(x)h(x)n}N+
\frac{e(x)\left(\partial_x a(x)-a(x)^2\partial_x h(x)n\right)}{(1+a(x)h(x)n)^2}N,\\
\psi_{2,1}(n,\hspace{-0.7mm}N\hspace{-0.7mm},\hspace{-0.3mm}x)
 & \hspace{-0.7mm}=\hspace{-0.5mm} &
\frac{\partial_{x^2}e(x)a(x)+2\partial_x e(x)\partial_x a(x)}{1\hspace{-0.5mm}+\hspace{-0.3mm}a(x)h(x)n}N\\
 & & \hspace{10.3mm} 
\frac{e(x)\left(\partial_{x^2}a(x) - a(x)^2\partial_{x^2}h(x)n-2a(x)\partial_x a(x)\partial_x h(x)n\right)}{(1\hspace{-0.5mm}+\hspace{-0.3mm}a(x)h(x)n)^2}N,\\
\psi_{2,2}(n,\hspace{-0.7mm}N\hspace{-0.7mm},\hspace{-0.3mm}x)
 & \hspace{-0.7mm}=\hspace{-0.5mm} &
-\frac{2\left(\partial_x a(x)h(x)+a(x)\partial_x h(x)\right)\partial_x e(x)a(x)}{(1\hspace{-0.5mm}+\hspace{-0.3mm}a(x)h(x)n)^2}N\\
 & & 
-\frac{2\left(\partial_x a(x)h(x)+a(x)\partial_x h(x)\right)e(x)\left(\partial_x a(x)-a(x)^2\partial_x h(x)n\right)}{(1\hspace{-0.5mm}+\hspace{-0.3mm}a(x)h(x)n)^3}N.
\end{eqnarray*}
\end{linenomath}
Taking the second mixed derivative of $F_3$ w.r.t.\hspace{-0.5mm} $x_{1,2}$, replacing replacing $n_3$ with $n\hspace{-0.3mm}-\hspace{-0.3mm}n_1\hspace{-0.3mm}-\hspace{-0.3mm}n_2$, simplifying, and collecting the monomial $n_1n_2$, one gets:
\begin{linenomath}
$$
\left.\partial_{x_1,x_2}\hspace{0.2mm}
F_3(n_1,n_2,n_3,N,x_1,x_2,x_3)\right|_{x_1=x_2=x_3=x} =
\psi_{1,1,1,1}(n,N,x)\hspace{0.2mm}n_1n_2,
$$
\end{linenomath}
with $\psi_{1,1,1,1}=\psi_{2,2}$ as expected from \eqref{eq:psi1111}.
Other relations between (same-order) $k$- and $k'$-index-$\phi$'s and $\psi$'s, $k'<k$, as well as to other consistency relations are checked in the Supplementary Material.

\newpage
{\small
\section*{Box 4}
\renewcommand{\theequation}{B4.\arabic{equation}}
\setcounter{equation}{0}  
Direct formulation of the single-species nonlinear fitness for the model of cannibalistic community described in \citet{Dercole_and_Rinaldi_02_TPB}.
The $g$-function is time-invariant and reads:
\begin{linenomath}
\begin{eqnarray}
\label{eq:gcan}
g_M(n_{1:M},x_{1:M},x') & \hspace{-0.5mm}:=\hspace{-0.2mm} &
e\,\Frac{a_0(x')\hspace{0.3mm}n_0\hspace{-0.7mm}+\hspace{-0.5mm}\makebox[5.5mm][l]{$\sum\limits_{j=1}^M$} a(x'\hspace{-0.5mm},x_j)\hspace{0.3mm}n_j}
{1\hspace{-0.7mm}+\hspace{-0.5mm}a_0(x')\hspace{0.3mm}h(x')\hspace{0.3mm}n_0\hspace{-0.5mm}+\hspace{-0.5mm}\makebox[5.5mm][l]{$\sum\limits_{j=1}^M$}a(x'\hspace{-0.5mm},x_j)\hspace{0.3mm}h(x')\hspace{0.3mm}n_j}\nonumber\\
 & & \hspace{-25mm}
-\sum\limits_{j=1}^M\Frac{a(x_j,x')\hspace{0.3mm}n_j}
{1\hspace{-0.7mm}+\hspace{-0.5mm}a_0(x_j)\hspace{0.3mm}h(x_j)\hspace{0.3mm}n_0\hspace{-0.7mm}+\hspace{-0.5mm}\makebox[5.5mm][l]{$\sum\limits_{k=1}^M$}a(x_j,x_k)\hspace{0.3mm}h(x_j)\hspace{0.3mm}n_k} - c\sum_{j=1}^Mn_j.
\end{eqnarray}
\end{linenomath}
For an individual with cannibalistic attitude $x'$, the growth rate $g_M$ describes reproduction in the first term, through the harvesting of a common resource $n_0$ and cannibalism toward the resident populations $n_j$, $j=1,\hspace{-0.6mm}...,\hspace{-0.3mm}M$, death due to cannibalistic attacks in the second term, and intra-specific competition for space and other limiting resources, at last
\citep[see][for a more detailed description of the model]{Dercole_and_Rinaldi_02_TPB}.

The $g$-function \eqref{eq:gcan} can be constructed with the procedure of Sect.~\ref{ssec:proc} in two steps ($L=2$) with time-invariant interacting kernels
\begin{linenomath}
\begin{subequations}
\begin{eqnarray}
e_n^{(1)}(x'\hspace{-0.7mm},\hspace{-0.3mm}x)\hspace{-0.3mm} &:=&
[\hspace{0.2mm}a(x'\hspace{-0.7mm},\hspace{-0.3mm}x),\hspace{0.3mm} a(x'\hspace{-0.7mm},\hspace{-0.3mm}x)h(x'),\hspace{-0.2mm} 1],\\
\label{eq:cane2}
e_n^{(2)}(x'\hspace{-0.7mm},\hspace{-0.3mm}x,E_n^{(1)}(x,x_{1:M}\hspace{-0.3mm},n_{1:M})\hspace{-0.3mm}) &:=&
\Frac{a(x,\hspace{-0.3mm}x')}{1\hspace{-0.5mm}+\hspace{-0.3mm}a_0(x)\hspace{0.3mm}h(x)\hspace{0.3mm}n_0\hspace{-0.5mm}+\hspace{-0.5mm}E_{n,2}^{(1)}(x,x_{1:M}\hspace{-0.3mm},n_{1:M})},
\end{eqnarray}
\end{subequations}
\end{linenomath}
and with the $g$-function
\begin{linenomath}
\begin{multline*}
g(x'\hspace{-0.7mm},\hspace{-0.5mm}E_n(x'\hspace{-0.7mm},\hspace{-0.3mm}x_{1:M}\hspace{-0.3mm},n_{1:M})\hspace{-0.3mm}) :=
e\hspace{0.3mm}\Frac{a_0(x')\hspace{0.3mm}n_0\hspace{-0.2mm}+\hspace{-0.5mm}E_{n,1}^{(1)}\hspace{-0.3mm}(x'\hspace{-0.7mm},\hspace{-0.3mm}x_{1:M}\hspace{-0.3mm},n_{1:M})}
{1\hspace{-0.5mm}+\hspace{-0.3mm}a_0(x')\hspace{0.3mm}h(x')\hspace{0.3mm}n_0\hspace{-0.5mm}+\hspace{-0.3mm}E_{n,2}^{(1)}(x'\hspace{-0.7mm},\hspace{-0.3mm}x_{1:M}\hspace{-0.3mm},n_{1:M})}\\
\hspace{-0.2mm}-\hspace{-0.5mm}E_n^{(2)}\hspace{-0.1mm}(x'\hspace{-0.7mm},\hspace{-0.3mm}x_{1:M}\hspace{-0.3mm},n_{1:M})
\hspace{-0.4mm}-\hspace{-0.4mm}cE_{n,3}^{(1)}\hspace{-0.1mm}(x'\hspace{-0.7mm},\hspace{-0.3mm}x_{1:M}\hspace{-0.3mm},n_{1:M}).
\end{multline*}
\end{linenomath}}

\vspace{-2mm}\noindent
Note that
$E_{n,2}^{(1)}(x'\hspace{-0.7mm},\hspace{-0.3mm}x_{1:M}\hspace{-0.3mm},n_{1:M})\hspace{-0.5mm}=\hspace{-0.8mm}
\makebox[10mm][l]{$\sum\nolimits_{j=1}^M$}a(x'\hspace{-0.5mm},x_j)\hspace{0.3mm}h(x')\hspace{0.3mm}n_j$
(the second component of $E_n^{(1)}$)
is evaluated with $x'\hspace{-0.5mm}=x$ in the denominator of $e_n^{(2)}$ and the integration of $e_n^{(2)}$ gives for $E_n^{(2)}\hspace{-0.1mm}(x'\hspace{-0.7mm},\hspace{-0.3mm}x_{1:M}\hspace{-0.3mm},n_{1:M})$ the sum in the second term of $g_M$ in \eqref{eq:gcan}.
Another way to see the two steps is noting that two indexes are necessary in the second term of $g_M$ to span the resident strategies: $j$ for the interactions of the $x'$-strategist (step 2) and $k$ for the concomitant interactions of the $x_j$-strategist (step 1).

The expressions of functions $\phi_{1,1}$, $\phi_{2,1}$, and $\phi_{2,2}$ are cumbersome to be reported here and are left in the Supplementary Material.

To show a case in which an interacting kernel (of level $l\ge 2$) of the focus species depends on the density of another interacting species (see eq.~\eqref{eq:E2}), let us now consider the dynamics of the common resource $n_0$, e.g. the logistic growth
\begin{linenomath}
\begin{eqnarray}
\label{eq:Fcan}
F_M(n_{1:M},n_0,x_{1:M}) & \hspace{-0.5mm}:=\hspace{-0.2mm} &
n_0\hspace{0.2mm}r\!\left(1-\Frac{n_0}{K}\right)-n_0\hspace{-0.5mm}
\sum\limits_{j=1}^M\Frac{a_0(x_j)\hspace{0.3mm}n_j}
{1\hspace{-0.7mm}+\hspace{-0.5mm}a_0(x_j)\hspace{0.3mm}h(x_j)\hspace{0.3mm}n_0\hspace{-0.7mm}+\hspace{-0.5mm}\makebox[5.5mm][l]{$\sum\limits_{k=1}^M$}a(x_j,x_k)\hspace{0.3mm}h(x_j)\hspace{0.3mm}n_k}.\nonumber\\
\end{eqnarray}
\end{linenomath}
It then becomes evident that $e_n^{(2)}$ in \eqref{eq:cane2} depends on $n_0$ (the left-hand side formally becomes $e_n^{(2)}(x'\hspace{-0.7mm},\hspace{-0.3mm}x,E_1(x,x_{1:M}\hspace{-0.3mm},n_{1:M}),n_0)$, $n_0$ now being a dynamic variable of model \eqref{eq:eco1} and not anymore a constant environmental parameter), and this is due to the fact that the intra-specific interaction described by $e_n^{(2)}$---predation---extends to the conspecifics strategies $x_1,x_2,\hspace{-0.6mm}...,x_M$ as well as to the common resource $n_0$.
The inter-specific interacting kernel here reads:
\begin{linenomath}
$$
e_N(x,E_n^{(1)}(x,x_{1:M}\hspace{-0.3mm},n_{1:M}),n_0):=
\Frac{a_0(x)}{1\hspace{-0.7mm}+\hspace{-0.5mm}a_0(x)\hspace{0.3mm}h(x)\hspace{0.3mm}n_0\hspace{-0.7mm}+
E_{n,2}^{(1)}(x,\hspace{-0.3mm}x_{1:M}\hspace{-0.3mm},n_{1:M})}
$$
\end{linenomath}
and gives the population growth rate \eqref{eq:Fcan} through the $F$-function
\begin{linenomath}
$$
F(E_N(x_{1:M},n_{1:M},n_0),n_0)=
n_0\hspace{0.2mm}r\!\left(1-\Frac{n_0}{K}\right)-n_0\hspace{0.2mm}
E_N(x_{1:M},n_{1:M},n_0).
$$
\end{linenomath}

\newpage

\begin{thebibliography}{22}
\expandafter\ifx\csname natexlab\endcsname\relax\def\natexlab#1{#1}\fi
\expandafter\ifx\csname url\endcsname\relax
  \def\url#1{\texttt{#1}}\fi
\expandafter\ifx\csname urlprefix\endcsname\relax\def\urlprefix{URL }\fi
\providecommand{\eprint}[2][]{\url{#2}}
\providecommand{\bibinfo}[2]{#2}
\ifx\xfnm\relax \def\xfnm[#1]{\unskip,\space#1}\fi
\bibitem[{Dercole(2003)}]{Dercole_03_JMB}
\bibinfo{author}{Dercole, F.}, \bibinfo{year}{2003}.
\newblock \bibinfo{title}{Remarks on branching-extinction evolutionary cycles}.
\newblock \bibinfo{journal}{J. Math. Biol.} \bibinfo{volume}{47},
  \bibinfo{pages}{569--580}.
\bibitem[{Dercole and Rinaldi(2002)}]{Dercole_and_Rinaldi_02_TPB}
\bibinfo{author}{Dercole, F.}, \bibinfo{author}{Rinaldi, S.},
  \bibinfo{year}{2002}.
\newblock \bibinfo{title}{Evolution of cannibalistic traits: Scenarios derived
  from adaptive dynamics}.
\newblock \bibinfo{journal}{Theor. Popul. Biol.} \bibinfo{volume}{62},
  \bibinfo{pages}{365--374}.
\bibitem[{Dercole and Rinaldi(2008)}]{Dercole_and_Rinaldi_08_PUP}
\bibinfo{author}{Dercole, F.}, \bibinfo{author}{Rinaldi, S.},
  \bibinfo{year}{2008}.
\newblock \bibinfo{title}{Analysis of Evolutionary Processes: The Adaptive
  Dynamics Approach and its Applications}.
\newblock \bibinfo{publisher}{Princeton University Press},
  \bibinfo{address}{Princeton, NJ}.
\bibitem[{Diekmann et~al.(2003)Diekmann, Gyllenberg and
  Metz}]{Diekmann_et_al_03_TPB}
\bibinfo{author}{Diekmann, O.}, \bibinfo{author}{Gyllenberg, M.},
  \bibinfo{author}{Metz, J.A.J.}, \bibinfo{year}{2003}.
\newblock \bibinfo{title}{Steady state analysis of structured population
  models}.
\newblock \bibinfo{journal}{Theor. Popul. Biol.} \bibinfo{volume}{63},
  \bibinfo{pages}{309--338}.
\bibitem[{Geritz(2005)}]{Geritz05a}
\bibinfo{author}{Geritz, S.A.H.}, \bibinfo{year}{2005}.
\newblock \bibinfo{title}{Resident-invader dynamics and the coexistence of
  similar strategies}.
\newblock \bibinfo{journal}{J. Math. Biol.} \bibinfo{volume}{50},
  \bibinfo{pages}{67--82}.
\bibitem[{Geritz et~al.(1998)Geritz, Kisdi, Mesz\'ena and Metz}]{Geritz98}
\bibinfo{author}{Geritz, S.A.H.}, \bibinfo{author}{Kisdi, E.},
  \bibinfo{author}{Mesz\'ena, G.}, \bibinfo{author}{Metz, J.A.J.},
  \bibinfo{year}{1998}.
\newblock \bibinfo{title}{Evolutionarily singular strategies and the adaptive
  growth and branching of the evolutionary tree}.
\newblock \bibinfo{journal}{Evol. Ecol.} \bibinfo{volume}{12},
  \bibinfo{pages}{35--57}.
\bibitem[{Geritz et~al.(1997)Geritz, Metz, Kisdi and Mesz\'ena}]{Geritz97}
\bibinfo{author}{Geritz, S.A.H.}, \bibinfo{author}{Metz, J.A.J.},
  \bibinfo{author}{Kisdi, E.}, \bibinfo{author}{Mesz\'ena, G.},
  \bibinfo{year}{1997}.
\newblock \bibinfo{title}{The dynamics of adaptation and evolutionary
  branching}.
\newblock \bibinfo{journal}{Phys. Rev. Lett.} \bibinfo{volume}{78},
  \bibinfo{pages}{2024--2027}.
\bibitem[{Greiner et~al.(1994)Greiner, Heesterbeek and
  Metz}]{Greiner_et_al_94_CAMQ}
\bibinfo{author}{Greiner, G.}, \bibinfo{author}{Heesterbeek, J.A.P.},
  \bibinfo{author}{Metz, J.A.J.}, \bibinfo{year}{1994}.
\newblock \bibinfo{title}{A singular perturbation theorem for evolution
  equations and time-scale arguments for structured population models}.
\newblock \bibinfo{journal}{Can. Appl. Math. Q.} \bibinfo{volume}{3},
  \bibinfo{pages}{435--459}.
\bibitem[{Gurney and Nisbet(1998)}]{Gurney98}
\bibinfo{author}{Gurney, W.S.C.}, \bibinfo{author}{Nisbet, R.M.},
  \bibinfo{year}{1998}.
\newblock \bibinfo{title}{Ecological Dynamics}.
\newblock \bibinfo{publisher}{Oxford University Press},
  \bibinfo{address}{Oxford, UK}.
\bibitem[{Gyllenberg and Metz(2001)}]{Metz_and_Gyllenberg_01_PRSB}
\bibinfo{author}{Gyllenberg, M.}, \bibinfo{author}{Metz, J.A.J.},
  \bibinfo{year}{2001}.
\newblock \bibinfo{title}{How should we define fitness in structured
  metapopulation models? {I}ncluding an application to the calculation of
  evolutionarily stable dispersal strategies}.
\newblock \bibinfo{journal}{Proc. Roy. Soc. Lond. {\rm B}}
  \bibinfo{volume}{268}, \bibinfo{pages}{499--508}.
\bibitem[{Hastings and Gross(2012)}]{Hastings_and_Gross_12_UCA}
\bibinfo{editor}{Hastings}, \bibinfo{editor}{Gross, L.} (Eds.),
  \bibinfo{year}{2012}.
\newblock \bibinfo{title}{Encyclopedia of Theoretical Ecology}.
  \bibinfo{publisher}{University of California Press},
  \bibinfo{address}{Berkeley}.
\bibitem[{Lotka(1920)}]{Lotka_20_JACS}
\bibinfo{author}{Lotka, A.J.}, \bibinfo{year}{1920}.
\newblock \bibinfo{title}{Undamped oscillations derived from the law of mass
  action}.
\newblock \bibinfo{journal}{J. Am. Chem. Soc.} \bibinfo{volume}{42},
  \bibinfo{pages}{1595--1599}.
\bibitem[{Lund(1965)}]{Lund_65_JCED}
\bibinfo{author}{Lund, E.W.}, \bibinfo{year}{1965}.
\newblock \bibinfo{title}{Guldberg and {W}aage and the law of mass action}.
\newblock \bibinfo{journal}{J. Chem. Educ.} \bibinfo{volume}{42},
  \bibinfo{pages}{548--550}.
\bibitem[{Mesz\'ena et~al.(2005)Mesz\'ena, Gyllenberg, Jacobs and
  Metz}]{Meszena05a}
\bibinfo{author}{Mesz\'ena, G.}, \bibinfo{author}{Gyllenberg, M.},
  \bibinfo{author}{Jacobs, F.J.}, \bibinfo{author}{Metz, J.A.J.},
  \bibinfo{year}{2005}.
\newblock \bibinfo{title}{Link between population dynamics and dynamics of
  {D}arwinian evolution}.
\newblock \bibinfo{journal}{Phys. Rev. Lett.} \bibinfo{volume}{95},
  \bibinfo{pages}{078105}.
\bibitem[{Mesz\'ena et~al.(2006)Mesz\'ena, Gyllenberg, P\'asztor and
  Metz}]{Meszena_et_al_06_TPB}
\bibinfo{author}{Mesz\'ena, G.}, \bibinfo{author}{Gyllenberg, M.},
  \bibinfo{author}{P\'asztor, L.}, \bibinfo{author}{Metz, J.A.J.},
  \bibinfo{year}{2006}.
\newblock \bibinfo{title}{Competitive exclusion and limiting similarity: A
  unified theory}.
\newblock \bibinfo{journal}{Theor. Popul. Biol.} \bibinfo{volume}{69},
  \bibinfo{pages}{68--87}.
\bibitem[{Metz et~al.(1992)Metz, Nisbet and Geritz}]{Metz92}
\bibinfo{author}{Metz, J.A.J.}, \bibinfo{author}{Nisbet, R.M.},
  \bibinfo{author}{Geritz, S.A.H.}, \bibinfo{year}{1992}.
\newblock \bibinfo{title}{How should we define fitness for general ecological
  scenarios?}
\newblock \bibinfo{journal}{Trends Ecol. Evol.} \bibinfo{volume}{7},
  \bibinfo{pages}{198--202}.
\bibitem[{Mylius and Diekmann(1995)}]{Mylius_and_Diekmann_95_OI}
\bibinfo{author}{Mylius, S.D.}, \bibinfo{author}{Diekmann, O.},
  \bibinfo{year}{1995}.
\newblock \bibinfo{title}{On evolutionarily stable life-histories, optimization
  and the need to be specific about density dependence}.
\newblock \bibinfo{journal}{Oikos} \bibinfo{volume}{74},
  \bibinfo{pages}{218--224}.
\bibitem[{Ruxton et~al.(1992)Ruxton, Gurney and {de
  Roos}}]{Ruxton_et_al_TPB_92}
\bibinfo{author}{Ruxton, G.D.}, \bibinfo{author}{Gurney, W.S.C.},
  \bibinfo{author}{{de Roos}, A.M.}, \bibinfo{year}{1992}.
\newblock \bibinfo{title}{Interference and generation cycles}.
\newblock \bibinfo{journal}{Theor. Popul. Biol.} \bibinfo{volume}{42},
  \bibinfo{pages}{235--253}.
\bibitem[{Thieme(2003)}]{Thieme_03_PUP}
\bibinfo{author}{Thieme, H.R.}, \bibinfo{year}{2003}.
\newblock \bibinfo{title}{Mathematics in Population Biology}.
\newblock \bibinfo{publisher}{Princeton University Press},
  \bibinfo{address}{Princeton, NJ}.
\bibitem[{Vincent and Brown(2005)}]{Vincent05}
\bibinfo{author}{Vincent, T.L.}, \bibinfo{author}{Brown, J.S.},
  \bibinfo{year}{2005}.
\newblock \bibinfo{title}{Evolutionary Game Theory, Natural Selection, and
  Darwinian Dynamics}.
\newblock \bibinfo{publisher}{Cambridge University Press},
  \bibinfo{address}{Cambridge, UK}.
\bibitem[{Volterra(1926)}]{Volterra26}
\bibinfo{author}{Volterra, V.}, \bibinfo{year}{1926}.
\newblock \bibinfo{title}{Variazioni e fluttuazioni del numero d'individui in
  specie animali conviventi}.
\newblock \bibinfo{journal}{Mem. Accad. Naz. Lincei} \bibinfo{volume}{2},
  \bibinfo{pages}{31--113}.
\newblock \bibinfo{note}{(in Italian)}.
\bibitem[{Waage and Gulberg(1864)}]{Waage_and_Gulberg_1864}
\bibinfo{author}{Waage, P.}, \bibinfo{author}{Gulberg, C.M.},
  \bibinfo{year}{1864}.
\newblock \bibinfo{title}{Studies concerning affinity}.
\newblock \bibinfo{journal}{{\it Forhandlinger: Videnskabs-Selskabet i
  Christiana}} \bibinfo{volume}{35}.
\newblock \bibinfo{note}{(in Norwegian; English translation by H. I. Abrash,
  {\it J. Chem. Educ.} 63, 1044--1047, 1986)}.

\end{thebibliography}



\end{document}